\documentclass[times,twocolumn]{aastex61}

\usepackage{graphicx}
\usepackage{subfigure}
\usepackage{natbib}
\usepackage{color}
\usepackage{tabularx}
\usepackage{bm}
\usepackage{threeparttable}

\newcommand{\nevents}{{217}} %Only 217 are unique events (i.e., some events have two OGLE numbers)
\newcommand{\thetae}{\theta_{\rm E}}
\newcommand{\pie}{\pi_{\rm E}}
\newcommand{\pievec}{\vec{\pi}_{\rm E}}

\newcommand{\tein}{t_{\rm E}}
\newcommand{\tzero}{t_0}
\newcommand{\uzero}{u_0}
\newcommand{\mcms}{{\it K2}C9-CFHT MCMS}
\newcommand{\uzearth}{u_{0,\earth}}
\newcommand{\uzkep}{u_{0,K2}}
\newcommand{\tzearth}{t_{0,\earth}}
\newcommand{\tzkep}{t_{0,K2}}
\newcommand{\tekep}{t_{{\rm E}, K2}}

\newcommand{\ktcn}{{\it K2}C9}
\newcommand{\kepler}{{\it Kepler}}
\newcommand{\kt}{{\it K2}}
\DeclareGraphicsExtensions{.pdf,.png,.jpg}

\shorttitle{\ktcn-CFHT Multi-color Microlensing Survey}
\shortauthors{Zang et al.}

\begin{document}

\title{Measurement of Source Star Colors with the \ktcn-CFHT Multi-color Microlensing Survey}

\author[0000-0001-6000-3463]{Weicheng Zang}
\affiliation{Physics Department and Tsinghua Centre for Astrophysics, Tsinghua University, Beijing 100084, China}
\affiliation{Department of Physics, Zhejiang University, Hangzhou, 310058, China}

\author[0000-0001-7506-5640]{Matthew T. Penny}
\affiliation{Department of Astronomy, The Ohio State University, 140 W. 18th Avenue, Columbus, OH 43210, USA}

\author{Wei Zhu}
\affiliation{Department of Astronomy, The Ohio State University, 140 W. 18th Avenue, Columbus, OH 43210, USA}
\affiliation{Canadian Institute for Theoretical Astrophysics, University of Toronto, 60 St George Street, Toronto, ON M5S 3H8, Canada}

\author{Shude Mao}
\affiliation{Physics Department and Tsinghua Centre for Astrophysics, Tsinghua University, Beijing 100084, China}
\affiliation{National Astronomical Observatories, Chinese Academy of Sciences, A20 Datun Rd., Chaoyang District, Beijing 100012, China}
\affiliation{Jodrell Bank Centre for Astrophysics, Alan Turing Building, University of Manchester, Manchester M13 9PL, UK}

\author{Pascal Fouqu\'e}
\affiliation{CFHT Corporation, 65-1238 Mamalahoa Hwy, Kamuela, Hawaii 96743, USA}
\affiliation{Universit\'e de Toulouse, UPS-OMP, IRAP, Toulouse, France}

\author{Andrzej Udalski}
\affiliation{Warsaw University Observatory, Al. Ujazdowskie 4, 00-478 Warszawa, Poland}

\author{Kyu-Ha Hwang}
\affiliation{Korea Astronomy and Space Science Institute, Daejon 34055, Korea}

\author{Tianshu Wang}
\affiliation{Physics Department and Tsinghua Centre for Astrophysics, Tsinghua University, Beijing 100084, China}

\author{Chelsea Huang}
\affiliation{Department of Physics and Kavli Institute for Astrophysics and Space Research, Massachusetts Institute of Technology, Cambridge, MA02139, USA}
\affiliation{Dunlap Institute for Astronomy and Astrophysics, University of Toronto, Toronto, ON M5S 3H4, Canada}
\affiliation{Centre of Planetary Science, University of Toronto, Scarborough Campus Physical \& Environmental Sciences, Toronto, M1C 1A4, Canada}

\author[0000-0001-9879-9313]{Tabetha. S. Boyajian}
\affiliation{Department of Physics and Astronomy, Louisiana State University, Baton Rouge, LA 70803 USA}

\author{Geert Barentsen}
\affiliation{Bay Area Environmental Research Institute, NASA Ames Research Center, P.O. Box 25, Moffett Field, CA 94035-0001}

%\collaboration{2016 CFHT-K2C9 Microlensing Survey}

%\collaboration{OGLE Collaboration}

\begin{abstract}\label{abs}
\kt\ Campaign 9 (\ktcn) was the first space-based microlensing parallax survey capable of measuring microlensing parallaxes of free-floating planet candidate microlensing events. Simultaneous to \ktcn\ observations we conducted the \ktcn\ Canada-France-Hawaii Telescope Multi-Color Microlensing Survey (\mcms) in order to measure the colors of microlensing source stars to improve the accuracy of \ktcn's parallax measurements. We describe the difference imaging photometry analysis of the \ktcn-CFHT MCMS observations, and present the project's first data release. This includes instrumental difference flux lightcurves of $\nevents$ microlensing events identified by other microlensing surveys, reference image photometry calibrated to PanSTARRS data release 1 photometry, and tools to convert between instrumental and calibrated flux scales. We derive accurate analytic transformations between the PanSTARRS bandpasses and the \kepler\ bandpass, as well as angular diameter-color relations in the PanSTARRS bandpasses. To demonstrate the use of our data set, we analyze ground-based and \kt\ data of a short timescale microlensing event, OGLE-2016-BLG-0795. We find the event has a timescale $t_{\rm E}=4.5 \pm 0.1$~days and microlens parallax $\pi_{\rm E}=0.12 \pm 0.03$ or $0.97 \pm 0.04$, subject to the standard satellite parallax degeneracy. We argue that the smaller value of the parallax is more likely, which implies that the lens is likely a stellar-mass object in the Galactic bulge as opposed to a super-Jupiter mass object in the Galactic disk.
\end{abstract}

\section{Introduction}\label{intro}

Gravitational microlensing opens a unique window to measuring the mass and population of free-floating planets (FFP). Typically, a Jupiter-mass lensing object has a short microlensing timescale $\sim1$ day. From an excess of short-timescale microlensing events in the MOA-II microlensing survey, \cite{2011Natur.473..349S} inferred the existence of a large population of unbound or wide-separation Jupiter-mass objects with $1.8_{-0.8}^{+1.7}$ per main sequence star. Such a result is in tension with theoretical expectations for wide-separation and ejected planets \citep{2016MNRAS.461L.107M,2017ApJ...834...46C} and surveys of young clusters \citep{2012ApJ...744....6S,2012ApJ...754...30P,2015ApJ...810..159M}. A more recent study of the timescale distribution of the OGLE-IV microlensing survey with a larger sample of events does not confirm the MOA result, placing an upper limit on the abundance of Jupiter-mass free-floating or wide-orbit planets of $0.25$ per main star at 95\% confidence \citep{Mroz2017}. While the population of Jupiter-mass objects appears to be small, \citet{2018AJ....155..121M} recently announced the discovery of the best free-floating planet candidate yet, with a most likely mass of the order of Neptune, though additional observations are required to confirm that it is indeed not bound to any star.

Proving that a short-timescale, apparently single point-mass microlensing event is caused by a free-floating planet requires both the measurement of the lens mass, and imaging observations to rule out the possibility of a host~\citep{Gould2016,Henderson2016b,Penny2017}. For a lensing object, the total mass is related to observables by
\begin{equation}
    M_{\rm L} = \frac{\thetae}{{\kappa}\pie} 
    \label{eq:mass}
\end{equation}
where $\kappa \equiv 4G/(c^2\mathrm{au}) = 8.144$ mas$/M_{\odot}$ is a constant, $\thetae$ is the angular Einstein radius, and $\pie$ is the microlensing parallax~\citep[e.g.,][]{Gould2000}. Measurement of $\thetae$ for non-luminous lenses requires measurement of finite source effects~\citep{1994ApJ...421L..75G,Shude1994,Nemiroff1994} and an estimate of the angular diameter of the source from its de-reddened color and magnitude~\citep[e.g.,][]{Kervella2008,Boyajian2014}. Even without a measurement of $\thetae$, the mass of lens can be strongly constrained if the parallax parameter $\pie$ is accurately measured \citep{1995ApJ...447...53H,Zhu2017}. In some cases, $\pie$ can be measured from the modulation in the lensing light curve with the orbital motion of the Earth around the Sun \citep{1992ApJ...392..442G, Alcock1995}. However, this method only works for microlensing events with long microlensing timescale $\tein\gtrsim$ year/$2\pi$ \citep[e.g.,][]{2017ApJ...845..129W}. In some rare cases, the parallax of a short-timescale event can be measured using terrestrial parallax \citep[e.g.,][]{Andy07244}. Thus the best way to measure the parallax of a short-timescale microlensing event is to observe it from two well-separated locations simultaneously \citep{1966MNRAS.134..315R,1994ApJ...421L..75G}. The feasibility of space-based microlensing parallax measurement has been demonstrated by \emph{Spitzer} microlensing programs \citep{2007ApJ...664..862D,2015ApJ...799..237U,2015ApJ...802...76Y,2015ApJ...805....8Z,2015ApJ...804...20C}. However, because scheduling \emph{Spitzer} observations requires a minimum 3 day turnaround time \citep{2015ApJ...799..237U} after discovery of the event, the \emph{Spitzer} microlensing program cannot measure the microlensing parallax of FFP candidate events, which have timescales ${\sim}1$~day.

From April 22 to July 2 of 2016, \kt's Campaign 9 (\ktcn) conducted the first space-based microlensing survey toward the Galactic bulge \citep{2016PASP..128l4401H}. The spatial separation between Earth and \kepler\ spacecraft enables the measurement of microlensing parallaxes $\pi_{E}$ for over ${\sim}100$-$200$ microlensing events. In particular, \ktcn's continuous, wide-field observations made it possible to measure parallax for FFP candidate events~\citep{Henderson2016b,Penny2017}. Thus \ktcn\ provided the first opportunity to uniquely measure the mass and the population of free-floating planets.

As reviewed by~\citet{2016PASP..128l4401H}, measurement of microlensing parallax requires measurement of the impact parameters $\uzearth$ and $\uzkep$ and times of maximum magnification $\tzearth$ and $\tzkep$, of the event as seen from Earth and \kepler, respectively.\footnote{In principle, the two events also have different timescales due to the different velocities of the observers. However, the difference in velocities is usually small compared to the relative velocities of the source and lens, so can be ignored.} These differ due to the displacement and relative velocity of the two observers. Unfortunately, because of \kepler's large $4''$ pixel scale, microlensing events detected via \kepler\ are highly blended. As a result, measurements of the $\uzkep$ and $\tekep$ are strongly degenerate with the source flux~\citep{Wozniak1997}, which will be a unique parameter for each observer, due to their different bandpasses. The degeneracy can be largely eliminated if it is possible to constrain the source flux in \kepler's bandpass. For many events observed from the ground with well sampled light curves, it is possible to accurately measure the source flux in the bandpass of a standard filter, but this differs from the source flux \kepler\ sees due to its broad $430$--$880$~nm bandpass \citep{2016PASP..128l4401H}. However, with measurements of the source flux in several standard filters spanning \kepler's bandpass it is possible to reconstruct the source flux that \kepler\ sees. 

In order to make the necessary measurements in several filters, with the Canada-France-Hawaii Telescope (CFHT) we conducted the 2016 \ktcn-CFHT Multi-color Microlensing Survey (\mcms) in the $g$-, $r$-, and $i$-bands, which together cover almost the entire \kepler\ bandpass.
\citet{2017PASP..129j4501Z} has shown that {\it K2} parallax uncertainties can be significantly reduced when a theoretical extinction dependent color-color relation is applied to derive the flux in \kepler's bandpass from photometry in the $V$ and $I$ bands. We therefore follow this method and derive relations between the \kepler\ bandpass and our filters. Our multi-color CFHT data also enable the measurement of the angular diameter of the microlensing sources through an angular diameter-color relation. This is crucial for mass measurement of FFP candidates when the finite-source effect arises \citep[e.g.][]{2018AJ....155..121M}, as it facilitates the conversion of the finite-source effect into measurement of the angular Einstein radius $\thetae$ of the lens, and thus their masses. 

Microlensing observations are usually conducted in the reddest optical bands due to large amounts of extinction towards the Galactic bulge. This makes color observations of sufficient depth and cadence for short-timscale free-floating planet candidates challenging from the small telescopes that usually conduct microlensing surveys. For this reason we designed and conducted the \mcms\ in order to maximize the scientific return of the \ktcn\ survey for the shortest timescale microlensing events. CFHT's large aperture, combined with the MegaCam instrument's wide field of view enabled deep imaging of the entire \ktcn\ superstamp in a short period of time to supplement the ground-based microlensing surveys' high-cadence observations in a single band and lower-cadence observations in bluer bands. CFHT's longitude complements those of other telescopes used for microlensing observations \citep[see Figure 9 of][]{2016PASP..128l4401H}, and Mauna Kea's excellent weather and observing conditions mean that CFHT can monitor parts of microlensing events that are poorly covered from the dedicated microlensing surveys.

In this paper we present the analysis and first public data release of the \mcms. This data release contains $g, r, i$ lightcurves of $\nevents$ previously identified microlensing events in our fields and calibrated photometry of our fields. The paper is structured as follows. In section~\ref{data}, we describe observations and data reduction procedures of the \mcms. We then derive the relation between ($K_p - r_{\rm PS1}$) vs. $(r - i)_{\rm PS1}$ in section~\ref{k2flux} and predict stellar angular diameters from PanSTARRS $g, r, i$ photometry in section~\ref{angular}. In section~\ref{OB160795}, we measure microlensing parallax for the short event OGLE-2016-BLG-0795 to test our method; and finally in section \ref{discussion}, we discuss the implications of our work.

\section{\mcms\ Data}\label{data}

\subsection{Survey Design}\label{survey}

\begin{figure}
    \centering
    \includegraphics[width=\columnwidth]{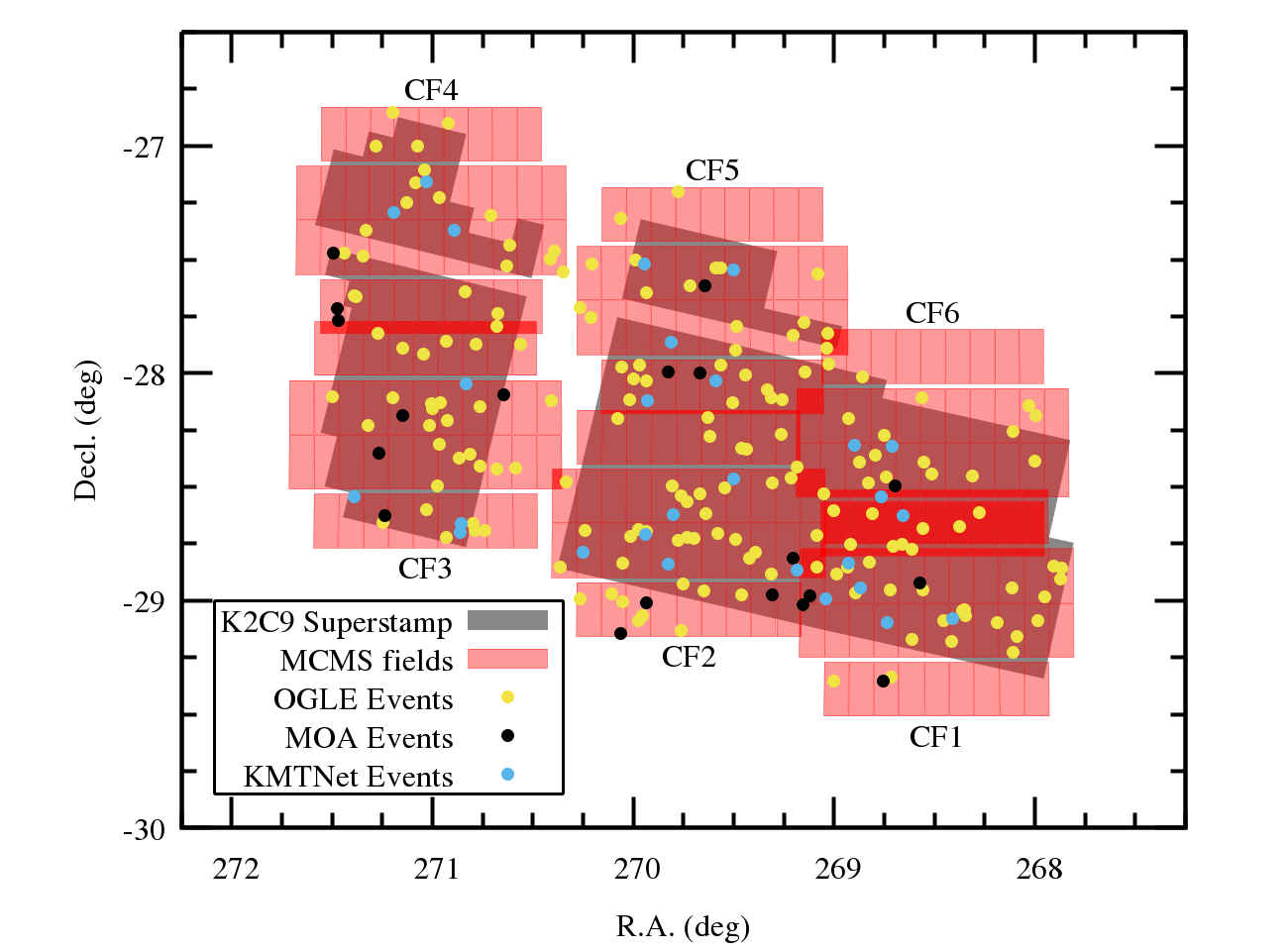}
    \caption{Map of the six \mcms\ fields, shown in pink transparency, which almost completely cover the ${\sim}3.7$~deg$^2$ \ktcn\ superstamp, shown in grey. Points show the $\nevents$ microlensing events for which we provide data in this release. Yellow dots show events alerted first by the OGLE Early Warning System~\citep{2003AcA....53..291U, 1994AcA....44..227U}, black dots show events alerted first by MOA \citep{2001MNRAS.327..868B}, and blue points show events only found by KMTNet~\citep{Kim2018b}.}
    \label{fig:fields}
\end{figure}

In 2016 we conducted the \mcms, a blind microlensing survey toward the Galactic bulge through the $g$-, $r$- and $i$-band filters using the MegaCam instrument on CFHT, a $0.94$~deg$^2$ imager built from 40 CCD chips with 0.187'' pixels, which are labeled ccd00 through ccd39 \citep{2003SPIE.4841...72B}. Six MegaCam pointings, labeled CF1 through CF6, were selected to cover the ${\sim}3.7 \mathrm{deg}^2$ K2C9 superstamp~\citep{Poleski2016, 2016PASP..128l4401H}, as shown in Figure~\ref{fig:fields}. Two to three observations per night in each filter were requested, with timing separated as widely as possible while the fields were observable. Observations were taken during 4 dark-time MegaCam runs, each lasting between 8 and 13 nights between HJD-2450000 = 7485 to 7575. Observations were collected on 42 nights. For $30$ of these at least two observations in each filter were taken, and the separation between observations in the same filter ranged from 0.5 to 4.5 hours.

Our aim was to measure the magnitude of the source star in each filter on a calibrated magnitude scale, in order to derive the source brightness in the \kepler\ bandpass and to measure the angular diameter of the source star. This proceeds through three steps. First, we constructed a high quality reference image from which to extract PSF photometry and to use to produce difference imaging analysis (DIA) photometry. Second, we produce a difference flux lightcurve using DIA image subtraction and photometry. Finally, this DIA lightcurve is tied to the flux scale of the reference image PSF photometry and then to an absolutely calibrated magnitude system.

We have not searched for microlensing events in our data. Instead, we rely on the OGLE Early Warning System~\citep{2003AcA....53..291U, 1994AcA....44..227U}, MOA microlensing alert system~\citep{2001MNRAS.327..868B}, and KMTNet event finder~\citep{Kim2018a, Kim2018b} to identify events. A search for new events in our data may be conducted in the future.

\subsection{Difference Imaging Photometry}\label{phot}

We produced differential flux lightcurves using a custom difference imaging analysis pipeline based on \texttt{ISIS} version 2.2 \citep{1998ApJ...503..325A,2000A&AS..144..363A}. Our processing began with images extracted from the standard bias subtracted and flat fielded images produced by the CFHT Elixir pipeline\footnote{http://www.cfht.hawaii.edu/Instruments/Elixir/}. Each $2048\times4612$~pixel (${\sim}6.4'\times14.4'$) chip and filter were processed separately, which proceeded through the following steps. 

 All images were registered to the same pixel grid as a single image selected by eye to have excellent seeing, low sky background level and a circular PSF, using the \texttt{IS3$\_$INTERP} interpolation routine and source extractor \footnote{http://verdis.phy.vanderbilt.edu/}~\citep{2012ApJ...761..123S,1996A&AS..117..393B}. A reference image was constructed by averaging between $1$ and $5$ images with a small and circular PSF, good seeing and a low background. PSF photometry of stars in the reference image was produced using the \citet{2012AJ....143...70A} implementation of \texttt{DoPHOT}~\citep{1993PASP..105.1342S}. These magnitudes were placed onto an approximately calibrated instrumental system, $(gri)_{\rm inst}$ using the Elixir photometric calibration parameters stored in the FITS header of one of the images used to construct the reference image. 
 
 For each microlensing event identified by the OGLE EWS \citep{2003AcA....53..291U, 1994AcA....44..227U}, MOA Alerts System \citep{2001MNRAS.327..868B}, and KMTNet event finder~\citep{2016JKAS...49...37K,Kim2018b}, we used \texttt{wcstools} \citep{Mink2011} to extract a $700~\times~700$~pixel sub-frame from each registered image and the reference image, centered on the Elixir-derived World Coordinate System position of the event. The reference image was convolved to match each target image and subtracted using the \texttt{ISIS} \texttt{sub} program. The precise centroid of the microlensing event was found using \texttt{ISIS}'s \texttt{detect} program operating on the variance image and assuming the event to be the object found within 5 pixels for OGLE and MOA events, 12 pixels for KMT events of the expected position of the microlensing target. Finally, difference photometry of the microlensing event was extracted from each difference image using the \texttt{ISIS} \texttt{phot} program with a PSF weighted aperture of 6 pixels in radius.

At the position of each star detected by \texttt{DoPHOT} in the full chip reference image, we also extracted photometry using the \texttt{ISIS}'s \texttt{phot} program with parameters identical to that used on the difference images. Using regression with iterative sigma clipping, we fit a linear slope to the instrumental counts (derived from the instrumental magnitudes) plotted against  the difference counts of bright stars with instrumental magnitudes in the ranges $14.5<i_{\rm inst}<18$, $15<r_{\rm inst}<18.5$ and $16.5<g_{\rm inst}<19.5$. Then the DIA lightcurve was multiplied by this slope to produce what we term the \emph{instrumental difference light curve} with the same magnitude zero point as our instrumental system. The uncertainty for this procedure was $\sim0.003$ mag in each filter. The light curves we provide in the data release are thus tied to the instrumental magnitude scale.

\subsection{Astrometric and Photometric Calibration}\label{PanSTARRS}

To place our photometry on an absolute system we cross-matched the instrumental DoPHOT catalog with PanSTARRS \citep{2016arXiv161205242M,2016arXiv161205243F} using the astrometric routines \texttt{grmatch} and \texttt{grtrans} of the Fitsh package \citep{2012MNRAS.421.1825P}. From this we derive an astrometric solution typically accurate to better than ${\sim}0.1$ pixel (0.0187 arcsec). From matched stars we sought to find a photometric transformation between our instrumental magnitudes in each filter $(g,r,i)_{\rm inst}$ and PanSTARRS DR1 \texttt{MeanPSFMag} calibrated magnitudes $(g,r,i)_{\rm PS1}$ of the form
\begin{equation}
    m_{\rm PS1}-m_{\rm inst} = a_{0,j} + a_1 C_{\rm inst} + f(x,y),
    \label{eqn:calib}
\end{equation}
where $C_{\rm inst}$ is a chosen instrumental color, $m$ is a chosen magnitude (e.g., $g$, $r$ or $i$), $a_{0,j}$ is the zeropoint of chip $j$, $a_{1}$ is the color term, and $f(x,y)$ is the variation of the zeropoint as a function of the pixel coordinates within a chip $x,y$. We investigated the use of higher order polynomial color terms, but found that the scatter in photometry of PanSTARRS-matched stars was too large to justify further terms. We found that the Elixir calibrated data contained systematic photometric offsets of up to a few percent as a function of position within a field that were common between fields. We found that a 2-d second order polynomial was sufficient to model these offsets and remove them.

Our fields are heavily crowded, which means that the PanSTARRS photometry has a relatively shallow limiting magnitude and contains a large fraction of outliers. For this reason we restricted our photometric calibration sample to a small range of instrumental magnitudes ($15.0<i_{\rm inst}<16.0$, $15.5<r_{\rm inst}<17.0$ and $16.5<g_{\rm inst}<18.5$; for the other filter appearing in the color term, only the bright limit was applied to remove stars near or above saturation. See \ref{fig:cmd} for color-magnitude diagram). We also required that each star had at least 20 detections (\texttt{nDetections}) in PanSTARRS, had reported uncertainties of less than $0.02$~mag in our instrumental photometry in both filters, and had \texttt{dophot} object code values of 1 in both filters. 

\begin{figure}
    \centering
    \includegraphics[width=\columnwidth]{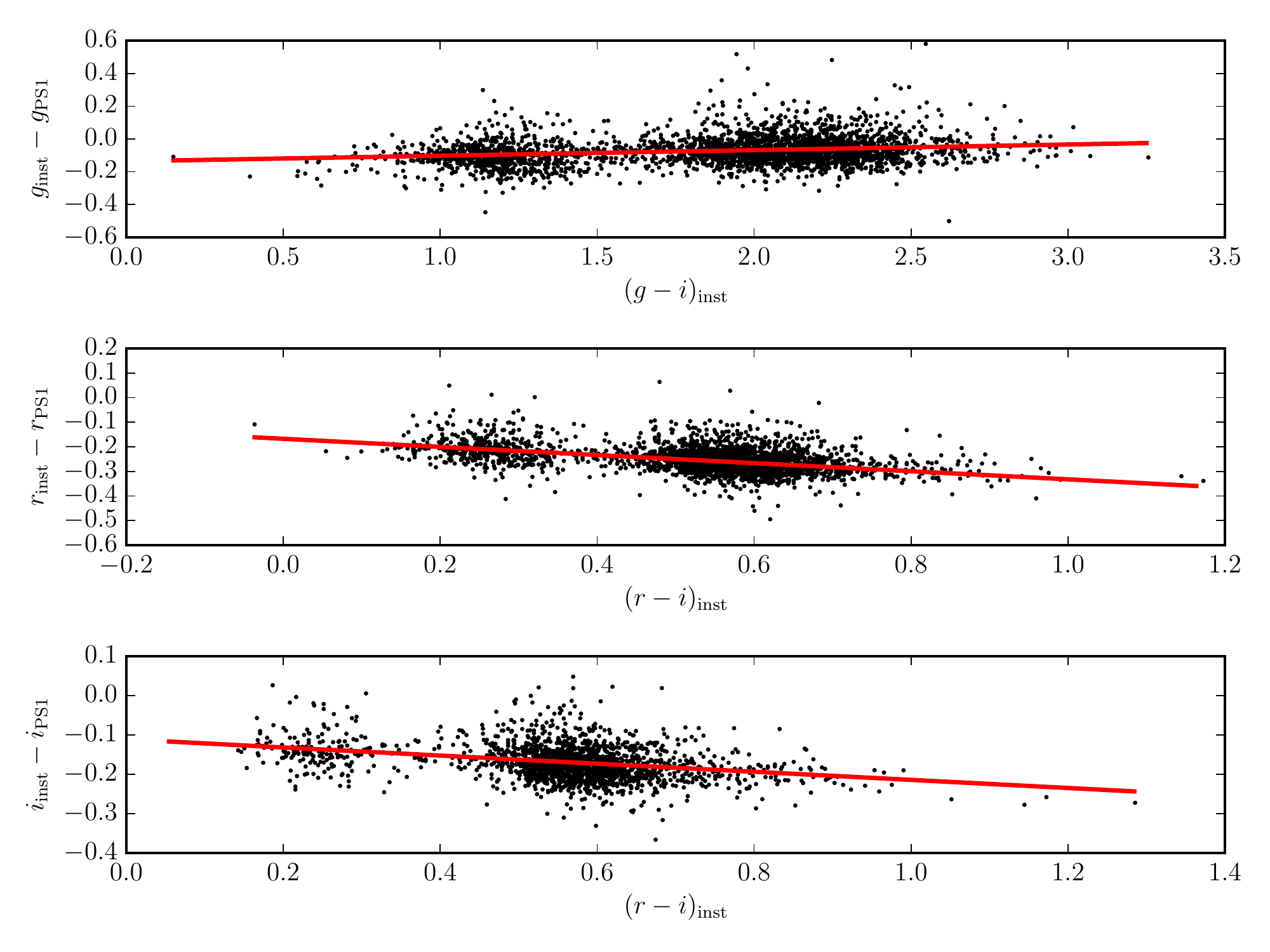}
    \caption{Difference between instrumental and PS1 photometry as a function of color after the subtraction of a preliminary zeropoint and flat field correction for data in field CF3 chip ccd13. The top, middle, and bottom panels show $g$, $r$, and $i$ data, respectively. Black points show photometry of stars, and the red line shows the result of a linear color term fit. Note that the lines were fit to data from the whole field covering a wider range of colors, and not just chip ccd13.}
    \label{fig:colorcal}
\end{figure}

\begin{figure}
    \includegraphics[width=\columnwidth]{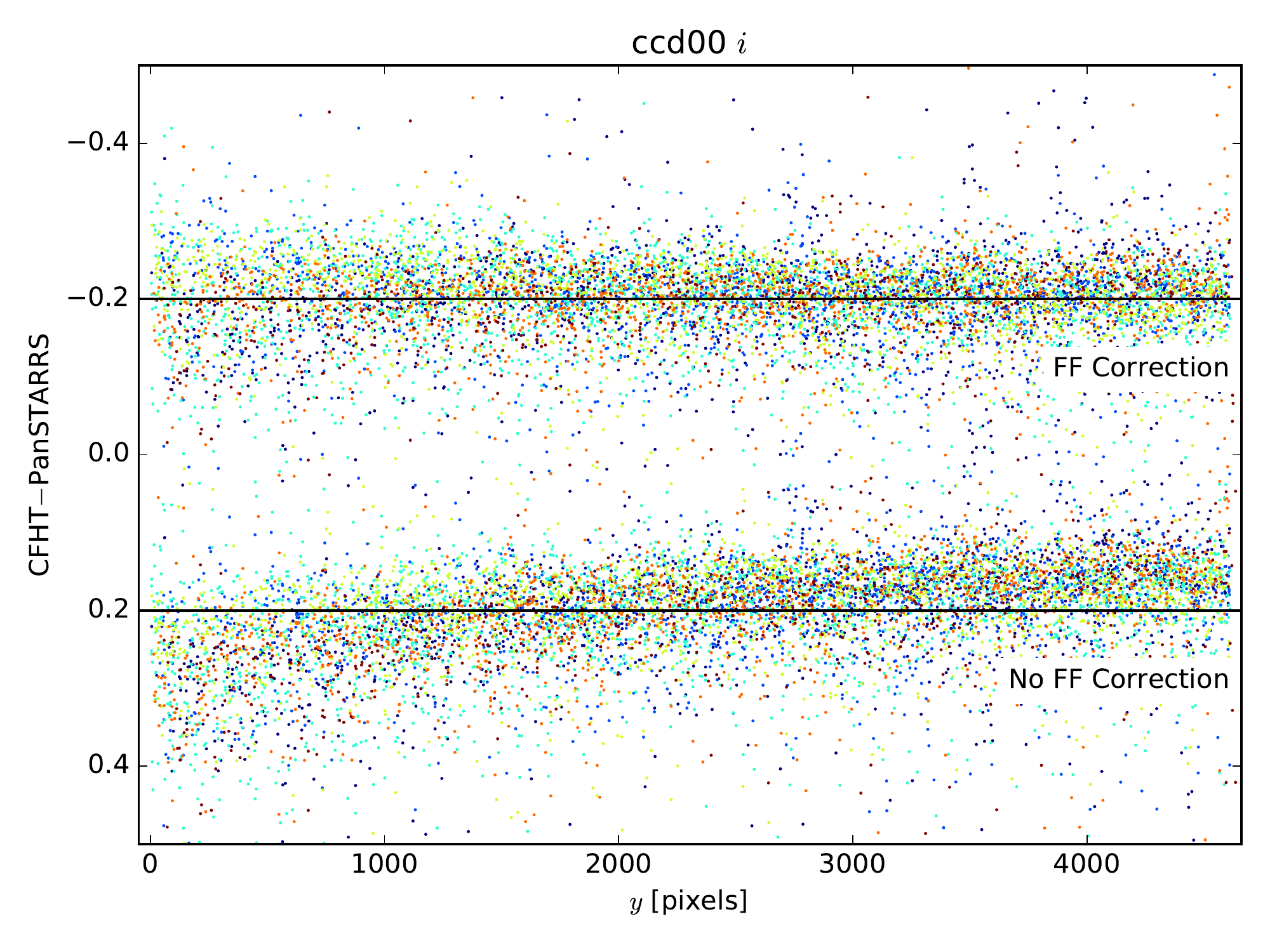}
    \caption{Difference between our calibrated $i$ magnitudes and PanSTARRS for chip ccd00 as a function of vertical pixel number before and after the flat field correction. Data from all fields is shown color-coded by field.}
    \label{fig:flatfield}
\end{figure}

Our fields cover a large range of reddening, so deriving a satisfactory transformation required fitting the color term $a_1$ across many different parts of the field to cover the entire range of colors and to minimize the effect of outliers. Therefore, for each of our six fields we performed the following three-step iterative procedure. We began by assuming an initial guess for the linear color term, and then fit for the individual chip zeropoints $a_{0,j}$. These zeropoints were subtracted and then we fit for the color terms. This process is repeated with updated coefficients until convergence is reached. Each fit employed iterative sigma clipping in 0.5 mag bins of magnitude or color. This process was conducted independently for each field to enable cross checks between the fields, and to allow for differences in atmospheric extinction between the fields. This provides the preliminary calibration. Flat field corrections were derived for each chip by combining the preliminary calibrated photometry from all fields and fitting a 2-d second-order polynomial to the photometry as a function of pixel coordinates. Finally, this polynomial was subtracted from the un-calibrated photometry and the first two steps were repeated again (fitting for the zeropoint and color term). Figure~\ref{fig:colorcal} shows an example of the color term fits, though only with data from one chip. Figure~\ref{fig:flatfield} shows the impact of applying the flat field correction for one of the worst affected chips.

\begin{figure}
\includegraphics[width=\columnwidth]{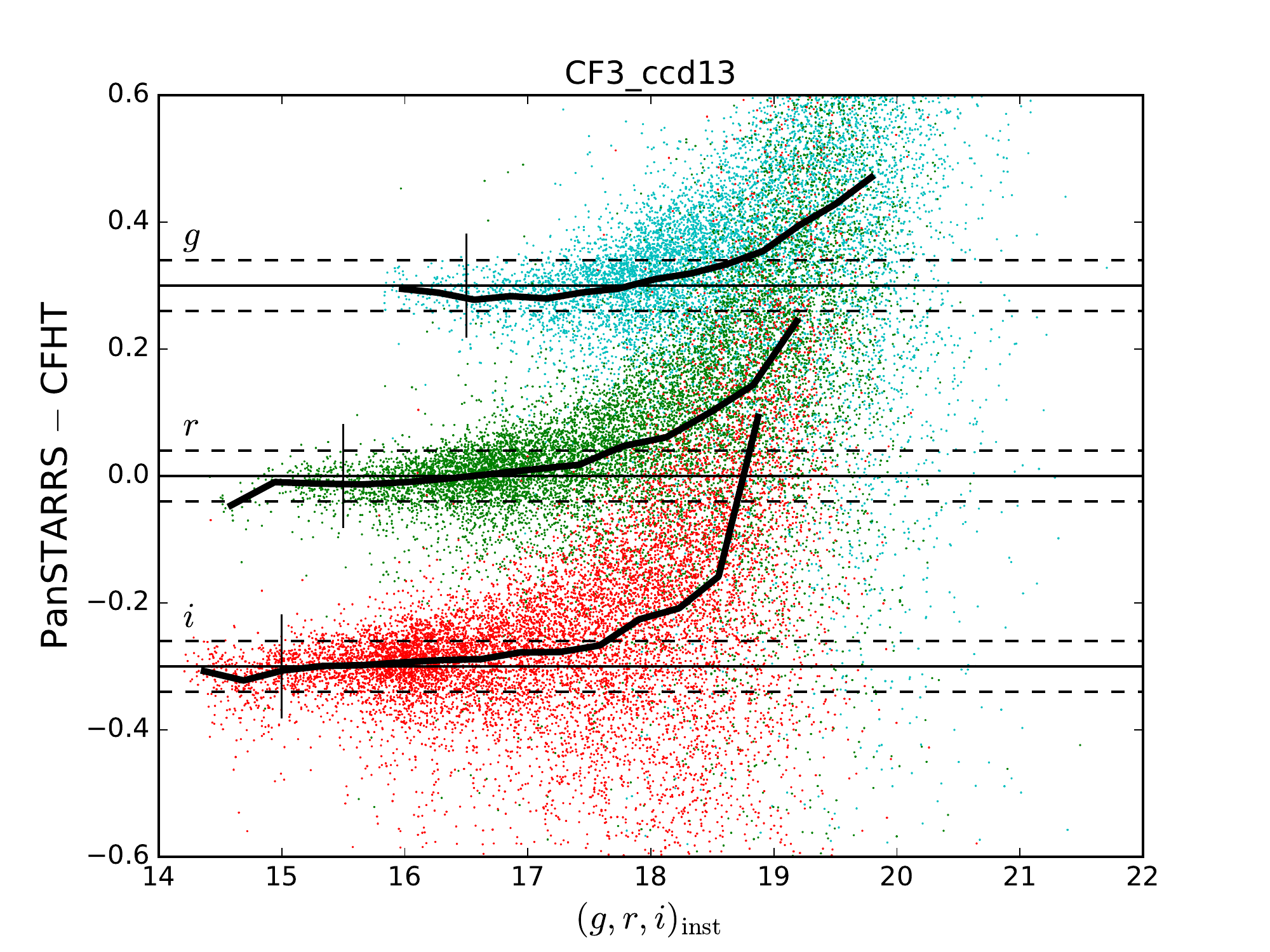}\\
\includegraphics[width=\columnwidth]{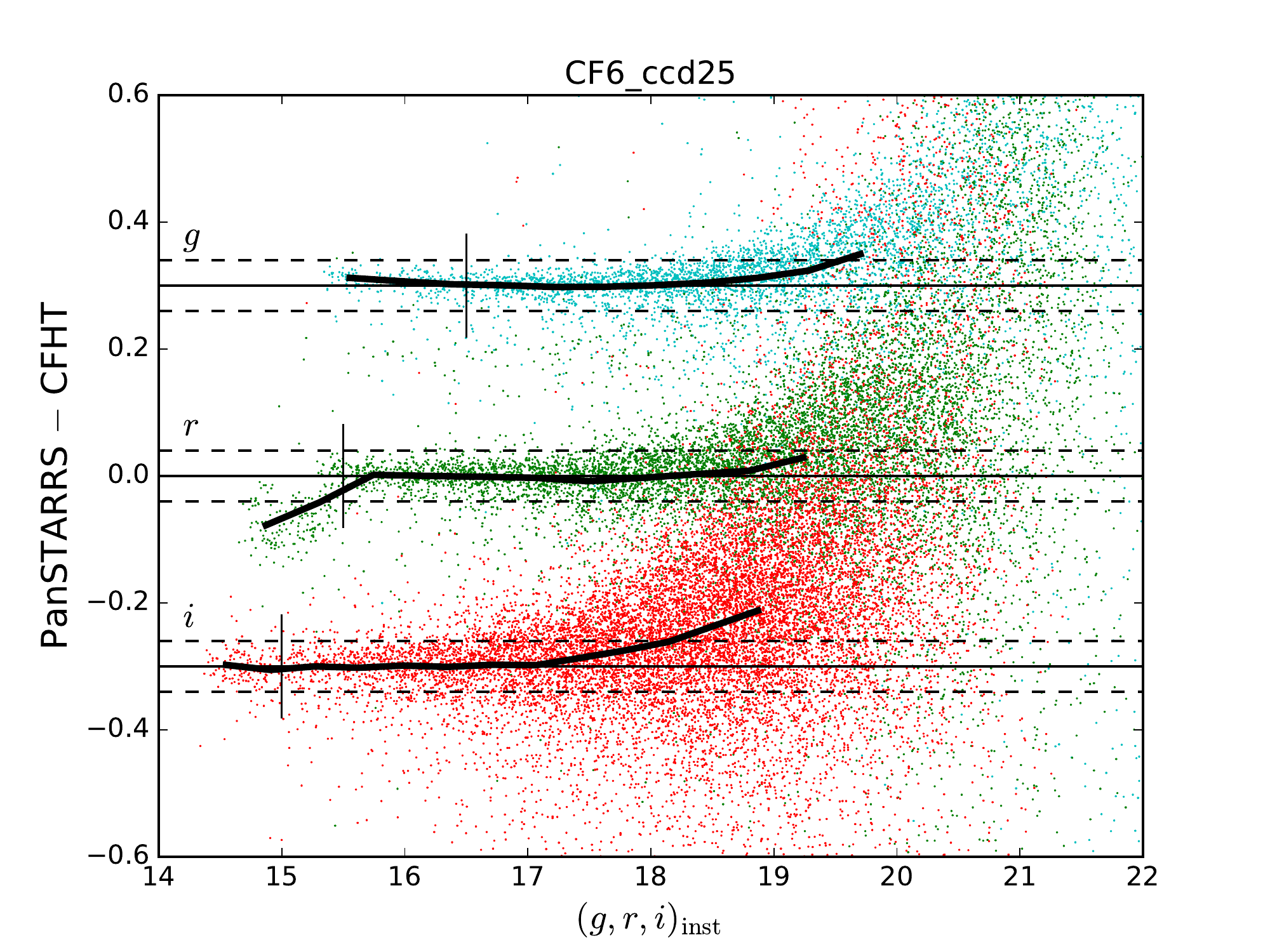}
\caption{Calibrated CFHT magnitudes compared to PanSTARRS catalog photometry for field CF3 chip ccd13, which contains the event OGLE-2016-BLG-0795, and field CF6 chip ccd25, a highly reddened area. Red, green and cyan points show the $g$, $r$ and $i$ photometry, respectively, for individual, cross-matched stars. The three filters are shifted vertically for clarity. Thick solid lines show the 3-sigma-clipped median of the points in bins, and thin solid lines show the exact-match line. Dashed lines are 0.04 mag either side of the exact-match line. Vertical lines show the bright limit of the magnitude range used for calibration, which is close to saturation.}
\label{fig:cal}
\end{figure}

 Figure~\ref{fig:cal} shows the difference between our calibrated photometry $(g,r,i)_{\rm PS1}$ and the PanSTARRS photometry for chip ccd13 in field CF3 (the field containing event OGLE-2016-BLG-0795 that we analyze later in the paper), and chip ccd25 in field CF6, chosen as an example of a field with strong extinction. The following features of the plots should be noted. Some of the photometry at the bright end of each filter might be affected by saturation. The exact magnitude at which this is an issue depends on the chip and the field. In every chip and every filter, we measure magnitudes that are brighter than PanSTARRS for fainter stars. This is almost certainly caused by crowding, as the onset of significant bias is approximately the bulge turnoff. Though we have not tracked down the exact cause of the bias, we expect that it is primarily a feature of crowded field photometry in PanSTARRS, rather than our own data, because our reference frames were taken with significantly better seeing than PanSTARRS on average. This bias makes it challenging to independently measure the accuracy of our transformations directly, in particular the zeropoint. 

Despite our removal of common-mode spatial trends in the photometry, there remain some chips with spatial offsets between our calibrated photometry and PanSTARRS. Some of this appears to be additional spatial variations in our own data that are not common between fields, while others appear to be structures in the PanSTARRS data potentially related to the PanSTARRS tiling. We have not attempted to correct for these remaining errors, instead we recommend that if users require a calibration to better than a few percent they inspect the PanSTARRS cross matched catalogs we provide and attempt their own empirical correction if it is needed.

\begin{figure}
\includegraphics[width=0.5\textwidth]{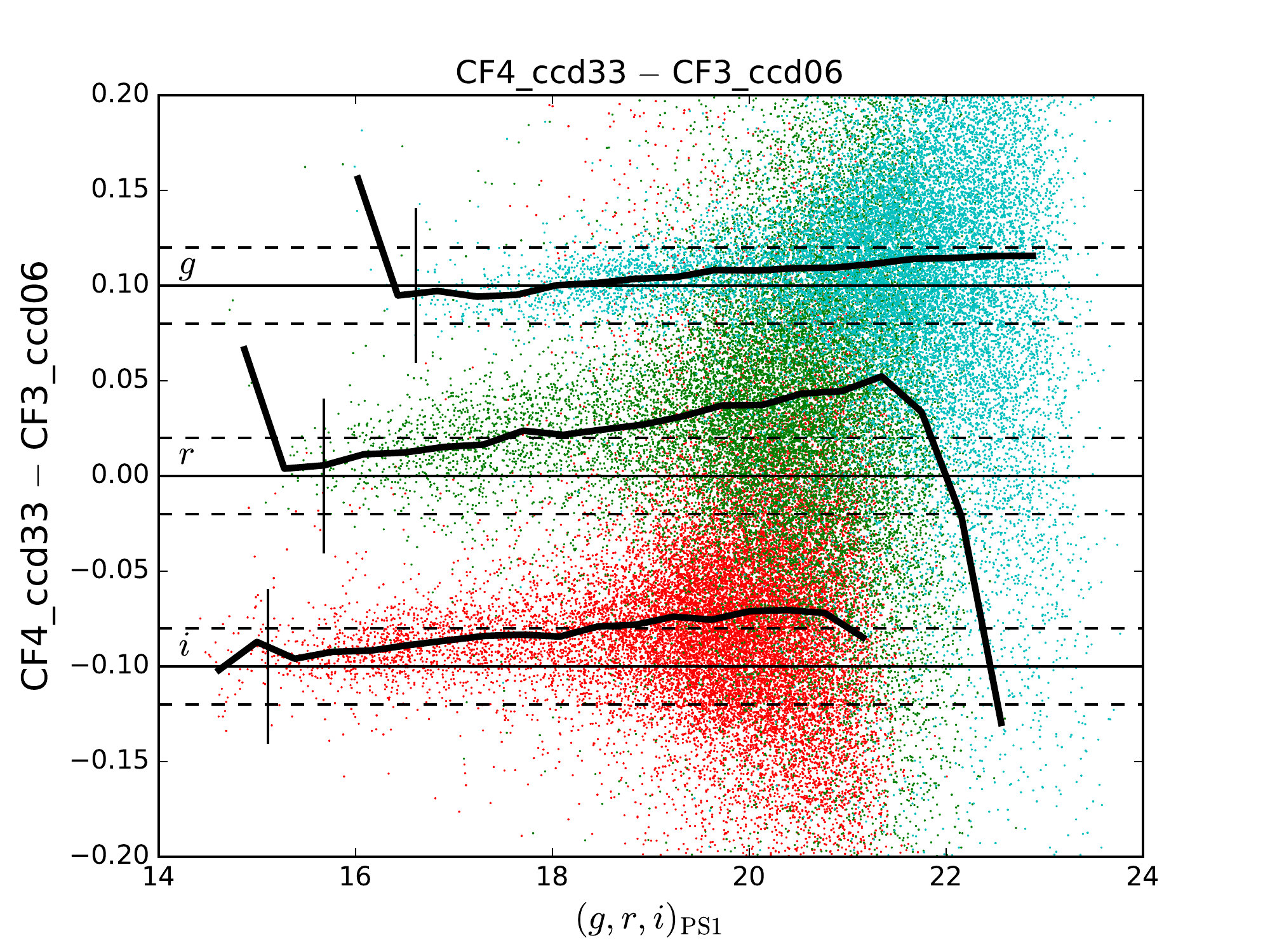}\\
\includegraphics[width=0.5\textwidth]{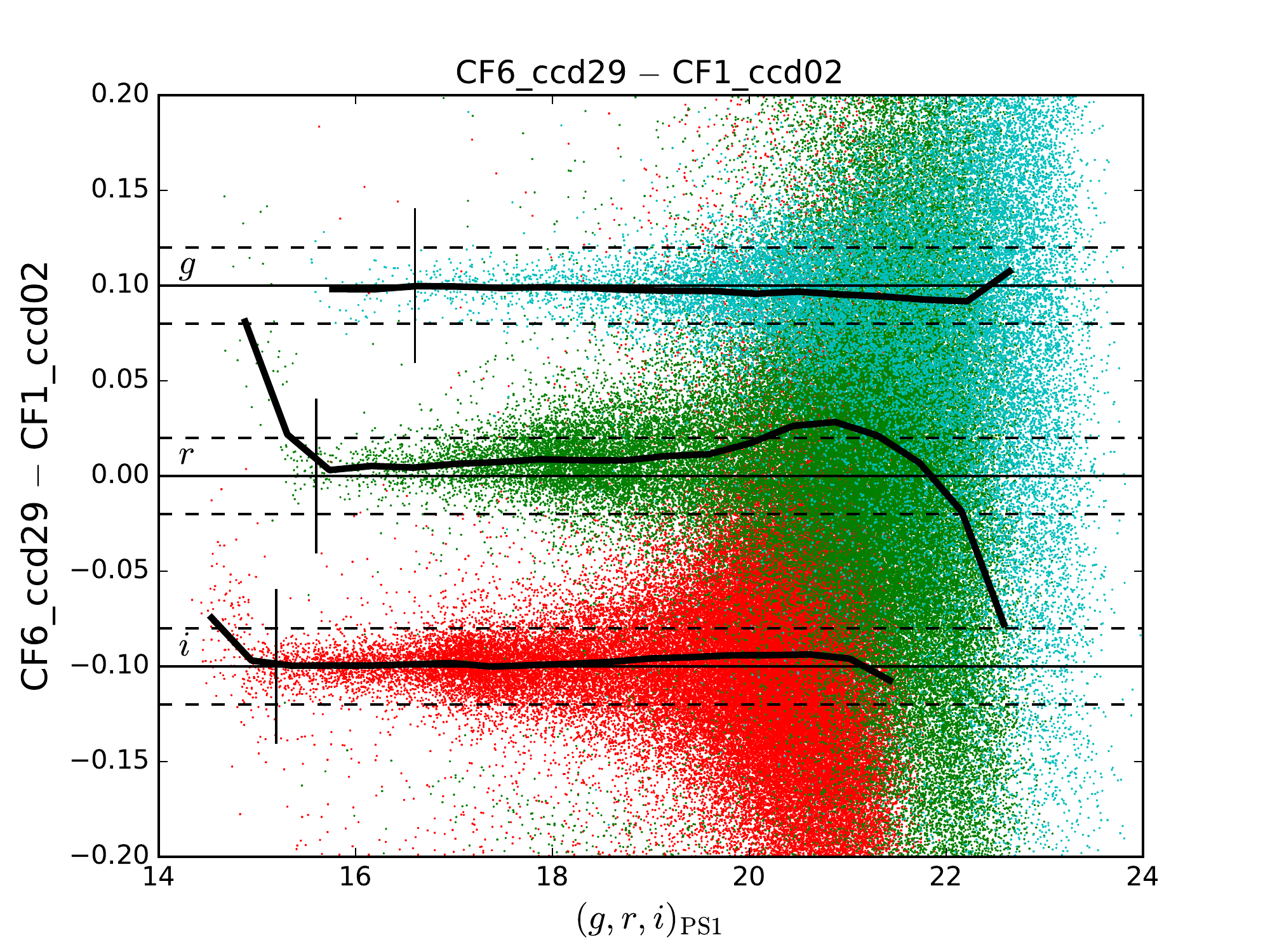}
\caption{Comparison of CFHT calibrated photometry in overlapping fields for internal consistency of our photometric calibration. The difference in our calibrated magnitudes between chips that overlap is plotted as a function of magnitude for chips in fields CF3 and CF4 (top) and in fields CF1 and CF6 (bottom). Dashed lines are 0.02 mag either side of the exact match line. Vertical lines are drawn slightly fainter than the approximate saturation magnitude. Note the factor of 3 smaller range of the $y$-axis scale compared to the plots in Figure~\ref{fig:cal}.} 
\label{fig:xmatch}
\end{figure}

\begin{table*}
    \centering
    \caption{Photometric calibration cross-check statistics}
    \begin{tabular}{lccccc}
\hline
Filter & Range & Median Offset & Max. Offset & Median Trend & Max. Trend \\
 & [mag] & [mag] & [mag] & [mag/mag] & [mag/mag] \\
\hline
$g$ & $17.5$-$21.0$ & $0.008$ & $0.027$ & $0.0030$ & $0.0128$ \\
$r$ & $16.5$-$19.0$ & $0.011$ & $0.024$ & $0.0031$ & $0.0087$ \\
$i$ & $16.0$-$18.0$ & $0.007$ & $0.032$ & $0.0029$ & $0.0158$ \\
\hline
\end{tabular}
    \label{tab:xmatch}
\end{table*}

While we can not easily check our photometric transformations against PanSTARRS, we can perform internal consistency checks where two of our fields overlapped. The largest region of overlap was between fields CF1 and CF6, though there is significant overlap between fields CF3 and CF4, and a few other smaller regions of overlap. For each pair of chips with significant overlap we cross matched sources and computed the median difference between calibrated magnitudes in a series of magnitude bins, which is shown for two  chip pairs in Figure~\ref{fig:xmatch}. There are systematic offsets in the differences, as well as trends in the differences with magnitude. To quantify these we measured the median offset and slope of a linear fit to the binned medians that are shown in Figure~\ref{fig:xmatch}. The means of the absolute offsets and absolute trend slopes over a restricted range of magnitudes are presented in Table~\ref{tab:xmatch} for each filter. From this we conclude that our photometry is internally consistent to ${\sim}0.01$~mag, and our photometry is linear to within ${\sim}0.003$ mag/mag, though there are some chip pairs with significantly worse disagreement. In one of these it appears to be caused by a faint saturation limit that fell within the calibration magnitude range in one of the chips. For most chips though, the trends are significantly smaller than the bias in the comparison of our photometry directly to PanSTARRS. While we have not tracked down a cause for the trends, we suspect they are not caused by crowding because they cover the entire magnitude range. To be conservative we adopt a systematic uncertainty in calibration for each of our bands of $0.02$~mag.

\subsection{Data Access}\label{access}

We make available photometry of the entire chip reference images, all microlensing event lightcurves, photometry for stars within 1 arcmin of each event, and calibration information in an institutional box\.com folder.\footnote{https://osu.box.com/v/CFHT-K2C9-MCMS-DR1} The lightcurve files include HJD timestampes, DIA photometry on the instrumental scale described in Section~\ref{phot}, airmass, DIMM seeing and a measure of the PSF full-width half maximum. Photometric catalogs include an accurate astrometric position, calibrated and instrumental magnitudes, \texttt{dophot} object flags, and pixel positions in each filter. In addition to this we provide a pdf-format summary page containing a color image of the chip, three plots of the calibrated full chip CMDs ($i$ versus $g-i$, $i$ versus $r-i$ and $r$ versus $g-r$), the PanSTARRS to CFHT comparison shown in Figure~\ref{fig:cal}, and plots summarizing each transformation equation fit. Finally, we also include a catalog of cross-matched stars between our data and PanSTARRS DR1.

The majority of time-series photometry data we have collected has very high quality. However, we have not cleaned our data for outliers. Most poor quality data arises from data taken with poor seeing or high airmass, allowing them to be cut using thresholds. Even doing this will leave some outliers that are caused by cosmic rays, bad CCD columns, or excess charge diffusion. These outliers will need to be expunged via sigma clipping, or some other method. Lightcurves that are affected by static CCD issues will likely be unusable because the telescope pointing does not vary significantly.

To convert from the difference flux units of the lightcurves to instrumental and calibrated magnitudes we provide a python script \texttt{calibrate\_flux.py} that can be downloaded from a github repository.\footnote{\url{https://github.com/mtpenny/cfht-microlensing}} This script takes as input combinations of $g, r$ and $i$ flux values and uncertainties, and returns magnitude data in the same format as the ps1calf files.

\section{Estimating \kepler\ Magnitudes with a Color-Color Relation}\label{k2flux}

\begin{figure}[htb] 
    \includegraphics[width=\columnwidth]{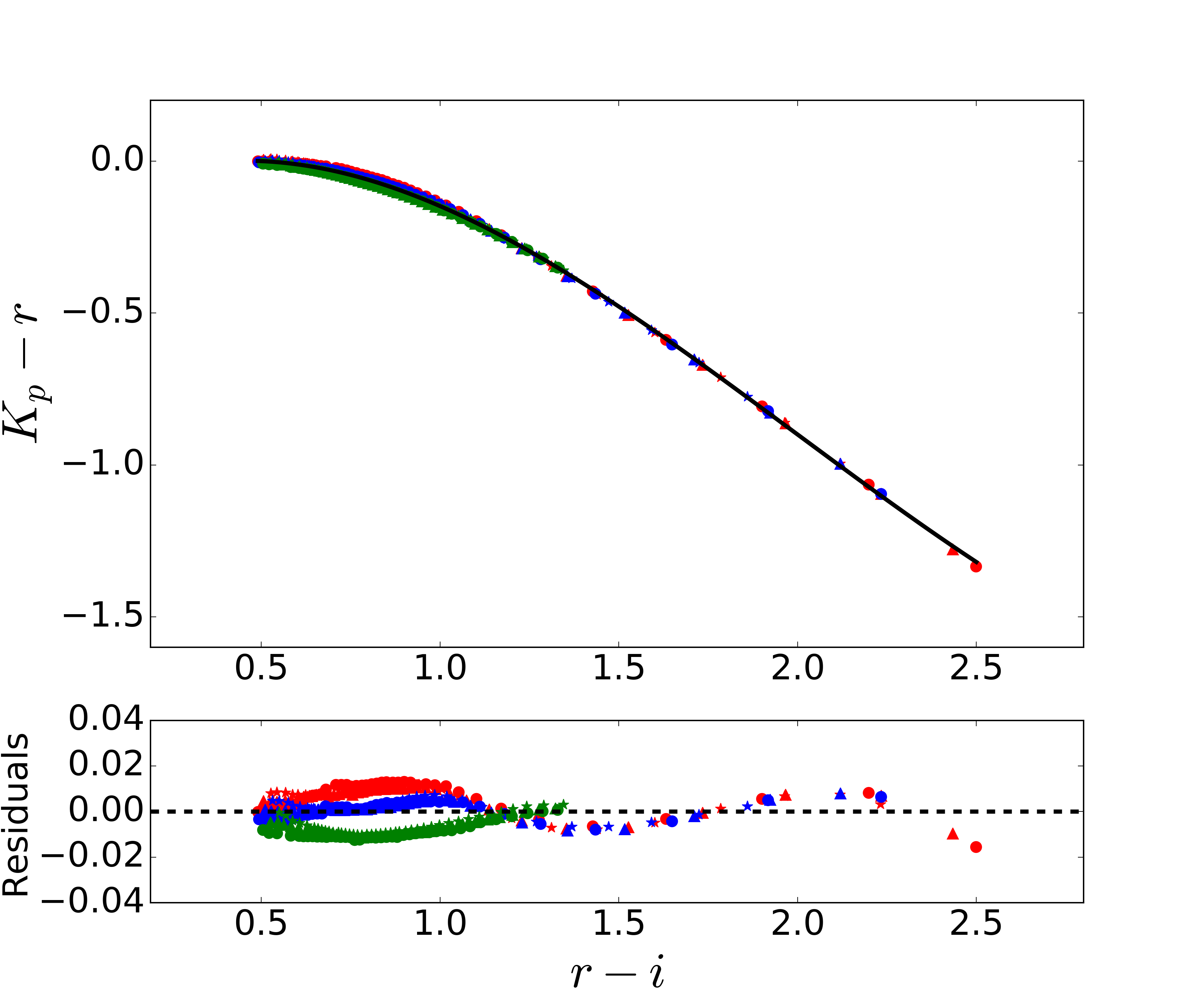}
    \caption{The $K_p - r~vs.~r - i$ color-color relations for extinction parameters $A_I = 2.0, R_I = 1.3$. \emph{Top panels}: Stellar colors are derived from the methods in Section~\ref{k2flux}, with different colors for different metallicities (green for [Fe/H]=-2, blue for [Fe/H]=0, and red for [Fe/H]=+0.5), shapes for surface gravity (circle for log$g$=2, triangle for log$g$=3, and asterisk for log$g$=4). \emph{Bottom panels}: Deviation from the best model for each data point.}
    \label{fig:kp}
\end{figure}

\begin{table*}[htb]
    \centering
    \caption{$K_p - r$ vs. $r - i$ Relation}
    \begin{threeparttable}
    \begin{tabular}{c c c c c c c c c}
    \hline
    \hline
    $A_I$ & $R_I$ & Range of $(r - i)$ & $b_0$ & $b_1$ & $b_2$ & $b_3$ & 1-$\sigma$ (mag) & Maximum Deviation (mag) \\
    \hline
    1.0 & 1.0 & 0.262--2.279 & -0.106 & 0.608 & -0.799 & 0.148 & 0.010 & 0.019 \\ 
    1.0 & 1.3 & 0.163--2.183 & -0.088 & 0.592 & -0.814 & 0.158 & 0.011 & 0.022 \\ 
    1.0 & 1.5 & 0.118--2.140 & -0.079 & 0.581 & -0.819 & 0.162 & 0.012 & 0.024 \\ 
    1.5 & 1.0 & 0.474--2.483 & -0.128 & 0.598 & -0.743 & 0.126 & 0.007 & 0.016 \\ 
    1.5 & 1.3 & 0.327--2.342 & -0.114 & 0.608 & -0.782 & 0.141 & 0.009 & 0.017 \\ 
    1.5 & 1.5 & 0.262--2.279 & -0.105 & 0.606 & -0.796 & 0.148 & 0.010 & 0.019 \\ 
    2.0 & 1.0 & 0.683--2.685 & -0.120 & 0.530 & -0.667 & 0.103 & 0.006 & 0.017 \\ 
    2.0 & 1.3 & 0.490--2.499 & -0.127 & 0.589 & -0.734 & 0.123 & 0.007 & 0.016 \\ 
    2.0 & 1.5 & 0.404--2.415 & -0.123 & 0.607 & -0.764 & 0.133 & 0.008 & 0.016 \\ 
    2.5 & 1.0 & 0.888--2.884 & -0.081 & 0.422 & -0.586 & 0.084 & 0.005 & 0.017 \\ 
    2.5 & 1.3 & 0.651--2.654 & -0.126 & 0.549 & -0.683 & 0.107 & 0.006 & 0.017 \\ 
    2.5 & 1.5 & 0.544--2.551 & -0.129 & 0.581 & -0.720 & 0.118 & 0.007 & 0.017 \\
    \hline
    \end{tabular}
    \begin{tablenotes}
        \item
        \item Note. --- Parameters for the cubic polynomial approximations of the $K_p - r_{\rm PS1}$ vs. $(r - i)_{\rm PS1}$ relation of Equation~\ref{eq:kp}, for different extinction parameters ($A_I~\&~R_I$). For each relation we have estimated both the 1-$\sigma$ uncertainty and the maximum deviation of the relation.
            \end{tablenotes}
        \end{threeparttable}
    \label{table:kp}
\end{table*}

Our goal is to estimate the flux of microlensed source in the \kepler\ bandpass. In this section we derive analytic relations between the PanSTARRS magnitudes and the \kepler\ magnitude, following the method described by \citet{2017PASP..129j4501Z}. For model spectra of stars with effective temperatures between 3400K and 8000K, and on a grid of surface gravity $\log g = (2, 3, 4)$ and metallicity [Fe/H]$= (-2.0, 0.0, +0.5)$, we calculated the stellar flux in the $K_p$ \citep{VanCleve2016} and $(g, r, i)_{\rm PS1}$ \citep{2012ApJ...750...99T} bandpasses using the PHOENIX synthetic stellar spectra \citep{2013A&A...553A...6H} multiplied by an extinction law that is linear in wavelength and defined by the total $I$-band extinction $A_I$ and the ratio of total to selective extinction $R_I=A_I/E(V-I)$, where $E(V-I)$ is the reddening in the $V-I$ color. We use extinctions and reddening in the OGLE-III Johnson-Cousins bandpasses, as these have been measured already by \citet{2013ApJ...769...88N}. After trying various color-color relations, we found that $K_p$ can be best estimated using the cubic polynomial
\begin{equation}
K_p - r_{\rm PS1} = \sum_{j=0}^3 b_j(A_I,R_I) (r-i)_{\rm PS1}^j,
\label{eq:kp}
\end{equation}
where $b_j(A_I,R_I)$ are the polynomial coefficients of order $j$. We fit for the polynomial coefficients of the color-color relation on a grid of $A_I=1.0,1.5,2.0,2.5$ and $R_I=1.0,1.3,1.5$,\footnote{The standard extinction law of $R_V=3.1$ corresponds to a value of $R_I=1.45$, but the majority of the bulge has an extinction law between $R_I=1.0$ and $R_I=1.4$~\citep{2013ApJ...769...88N}} and the results are presented in Table~\ref{table:kp}. The exact form of the extinction law that we chose has only a small effect on the relations. The synthetic colors and the fitted polynomial are shown in Figure~\ref{fig:kp} for $(A_I,R_I)=(2.0,1.3)$. The polynomial approximation is accurate to within ${\sim}0.01$~mag. For a typical microlensing event where daily CFHT data has a good coverage, the uncertainty on $r_{PS1} - i_{PS1}$ is $\sim0.03$~mag. Then the uncertainty on $K_p - r_{PS1}$ is $0.02\sim0.03$~mag for different extinction parameters.

\section{Stellar Angular Diameters Estimates from P\lowercase{an}STARRS Photometry}\label{angular}

If finite source effects are measured in the lightcurve of a microlensing event, they can be used to estimate the angular Einstein radius and relative source-lens proper motion~\citep{1994ApJ...421L..75G,Shude1994,Nemiroff1994}. When such a measurement is combined with a microlensing parallax, it is possible to solve for the mass and distance of the lens (up to any parallax degeneracy). Finite source effects are parameterized by the parameter $t_*$, the time taken to cross the radius of the source. The ratio $t_*/\tein$ is equal to the ratio of the source star angular radius $\theta_*$ to the angular Einstein radius $\theta_{\rm E}$. The angular diameter of the source can be estimated from the dereddened source color and magnitude through an angular diameter-color relation of the form~\citep[e.g.][]{Kervella2008}
\begin{equation}
\log 2\theta_* = \log \theta_{m=0} - 0.2 m,
\label{eq:ad}
\end{equation}
where $m$ is a dereddened apparent magnitude in a chosen filter, and $\theta_{m=0}$ is the angular diameter of a star with zero apparent magnitude in the chosen filter. $\log \theta_{m=0}$ is a function of color that can be approximated by a polynomial.

Angular diameter-color relations have not been derived previously for the PanSTARRS filters. So we do it here, choosing $i_{\rm PS1}$ for the magnitude, and deriving relations for both $(g-i)_{\rm PS1}$ and $(r-i)_{\rm PS1}$. We follow the methodology, and use the data set of \citet{Boyajian2014}, to fit for the polynomial coefficients of the relation. We use the sample of limb-darkening corrected, interferometric angular diameters and Sloan photometry compiled by \citet{2012ApJ...757..112B,2013ApJ...771...40B}. All stars are required to have a diameter error $<3$\%. The $g, r, i$ photometry in the Sloan bandpasses was synthetically computed using two different methods respectively \citep{2008PASP..120.1128O,2010PASP..122.1437P}. We reject any star where the synthetic magnitudes disagree by more than $0.15$ magnitude in any one band, and take the average of the two magnitudes. We transform the Sloan magnitudes to the PanSTARRS $(g, r, i)_{\rm PS1}$ system using the transformations of \citet{2016ApJ...822...66F}. We assume the magnitude uncertainties are $0.06, 0.04$ and $0.04$~mag in $g$, $r$ and $i$, respectively, based on those reported by \citet{2010PASP..122.1437P}, and assume the color uncertainty is the larger of the component uncertainties as we expect the uncertainties on each magnitude to be strongly correlated. 

\begin{figure*}[t!]
    \centering
    \begin{subfigure}
    \centering
    \includegraphics[width=\columnwidth]{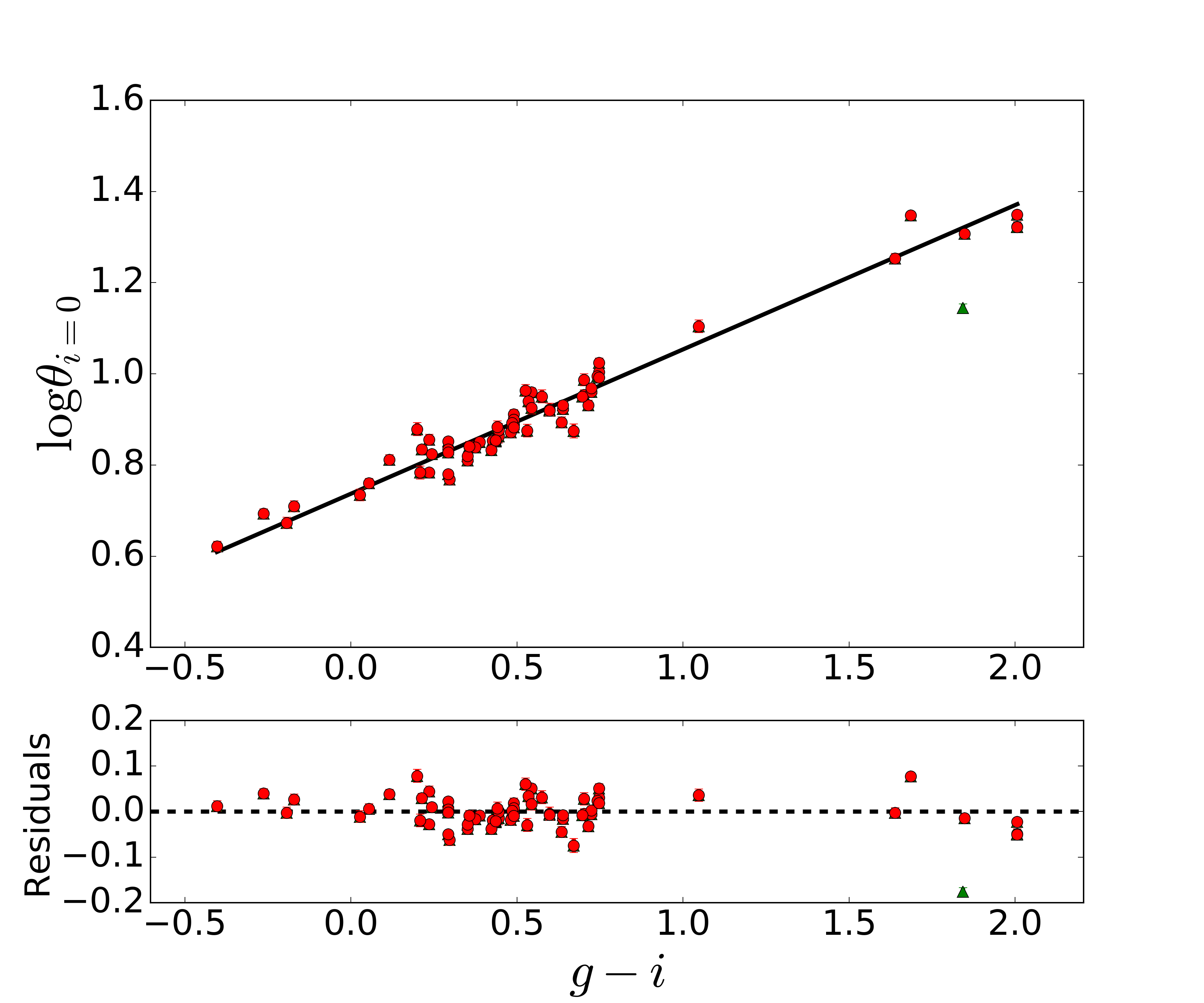}
    \end{subfigure}
    ~
    \begin{subfigure}
    \centering
    \includegraphics[width=\columnwidth]{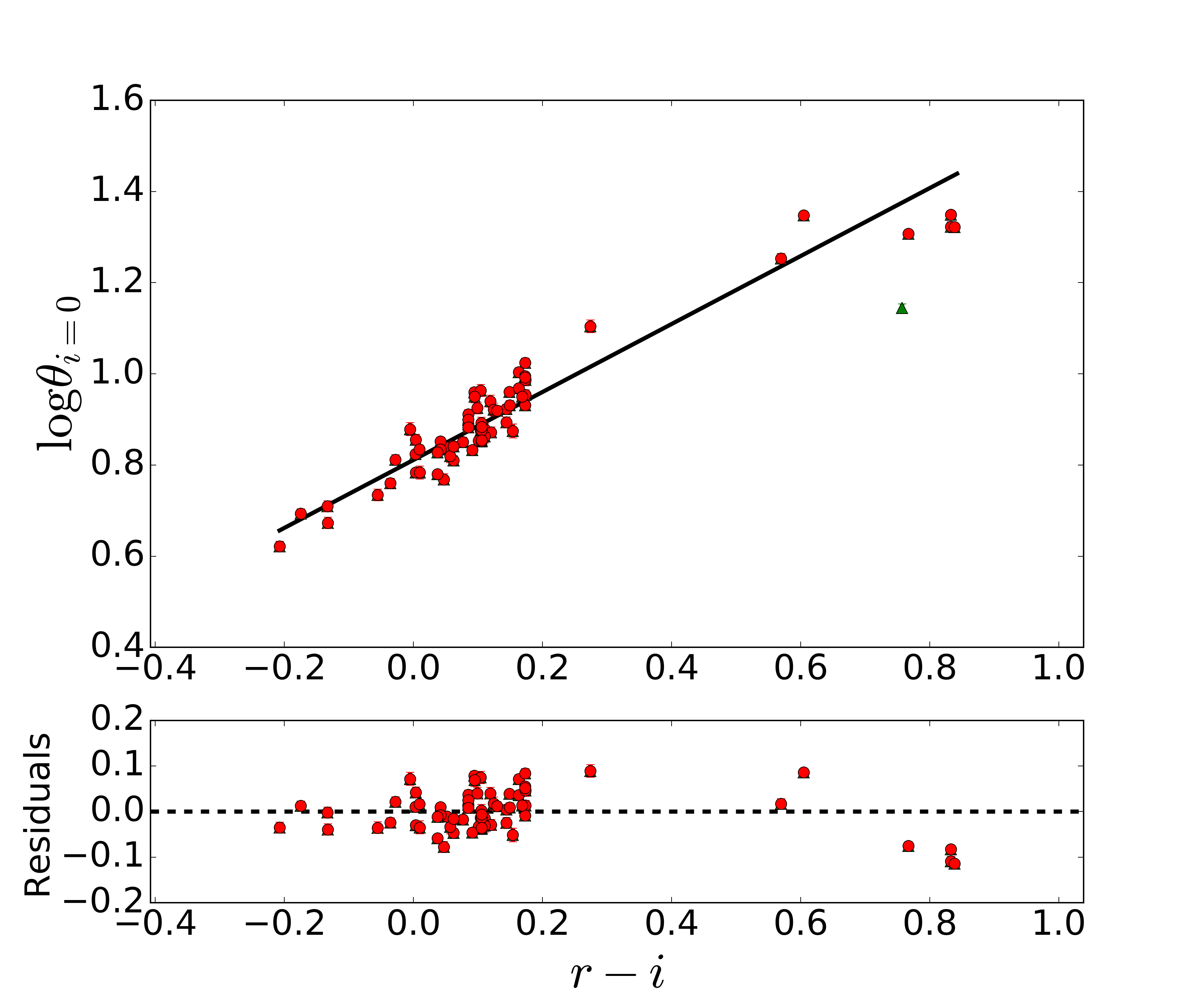}
    \end{subfigure}

    \caption{Angular diameter-color relations for the PanSTARRS $i$ magnitudes and $(g-i)_{\rm PS1}$ (left panels) and $(r-i)_{\rm PS1}$ (right panels) colors. Measurements are shown as points with error bars (red dots are used in the fit, and green dots were excluded by iterative sigma clipping), and the best linear fits are shown with the black lines. The bottom panels show the residuals to the fit.}
    \label{fig:theta}
\end{figure*}

\begin{table*}[tb]
    \centering
    \caption{Angular Diameter-Color Relations}
    \begin{threeparttable}
    \begin{tabular}{c c c c c c c c c}
    \hline
    \hline
    Color & No. of Points & Range & $c_0$ & $c_1$ & $X_{\rm p}$ & RMS (dex) & Pred. Frac. Uncertainty \\
    \hline
    $(g-i)_{\rm PS1}$ & 66 & -0.403--2.006 & $0.920\pm0.004$ & $0.316\pm0.008$ & 0.576 & 0.031 & 0.035--0.060 \\ 
    $(r-i)_{\rm PS1}$ & 66 & -0.207--0.839 & $0.920\pm0.006$ & $0.745\pm0.028$ & 0.146 & 0.046 & 0.075--0.119 \\
    \hline
    $(V-I)_{\rm C}$ & 83 & -0.050--1.740 & $0.542\pm0.006$ & $0.391\pm0.006$ & 0 & 0.028 & 0.043--0.065 \\
    \hline
    \end{tabular}
    \begin{tablenotes}
        \item
        \item Note. --- Fitted parameters for the linear relations between angular diameter and PanSTARRS colors. In addition to the parameters, listed are the number of stars used in the fit, the range of color for which the fit is valid, the RMS of data about the fitted relation, and the range of propagated fractional uncertainties for the angular diameter that would be derived using Equation~\ref{eq:ad}, assuming 0.03 mag errors in each band, which will usually be conservative. For comparison we also show the $(V - I)_{\rm C}$ relation from \cite{Adams2018}, which is more accurate thanks to more accurate photometry of the bright stars with interferometric angular diameters.
            \end{tablenotes}
        \end{threeparttable}
    \label{table:angular}
\end{table*}

Using regression with iterative sigma clipping, we find that $\log\theta_{i_{\rm PS1}=0}$ are well described as linear functions in both $(g-i)_{\rm PS1}$  and $(r-i)_{\rm PS1}$. We therefore fit for the parameters of the equation
\begin{equation}
    \log \theta_{i_{\rm PS1}=0} = c_0 + c_1 (X-X_{\rm p}),
    \label{eq:zmld}
\end{equation}
where $X$ is a chosen color, and $X_{\rm p}$ is a pivot color chosen to minimize the covariance between the parameters. The data and best fit lines are shown in Figure~\ref{fig:theta}, and the pivot colors and coefficients of the fit are given in Table~\ref{table:angular}. The relation for $(g - i)_{\rm PS1}$ is more precise, with estimated fractional uncertainties of $3.5$--$6.0$\% on $\theta_{i_{\rm PS1}}$ for an input . The $(r - i)_{\rm PS1}$ relation has worse precision with uncertainties in the range of $7.5$--$11.9$\%. However, in cases of significant extinction, the $(r - i)_{\rm PS1}$ relation can be more accurate. We also caution that the $(g-i)$ relation may be metallicity dependent at the level of ${\sim}0.08$ per dex in $\log\theta_*$, \citet{Boyajian2014} having statistically significant evidence for a metallicity dependence in the $g-r$ color-angular diameter relation. Our $(g-i)_{\rm PS1}$ relation may therefore have a significant, undiagnosed systematic error that makes the $(r-i)_{\rm PS1}$ relation relatively more competitive. Finally, we note that the largest source of error in deriving our relations was the available $ugriz$-system photometry for the bright stars used, and not the angular diameters.

\section{A short event: OGLE-2016-BLG-0795}\label{OB160795}

In this section, we demonstrate the use of our multi-color photometry by analyzing a short timescale microlensing event, OGLE-2016-BLG-0795, that occurred during \ktcn.

\subsection{Ground-based Observations}\label{ground}

OGLE-2016-BLG-0795 was firstly discovered by the OGLE collaboration using its 1.3 m Warsaw Telescope at the Las Campanas Observatory in Chile \citep{2015AcA....65....1U}, and alerted by the OGLE Early Warning System \citep{2003AcA....53..291U, 1994AcA....44..227U}. The event was located at equatorial coordinates $(\alpha, \delta)_{\rm J2000} = (18^{\rm h}04{\arcmin}00{\farcs}26, −28^{\circ}09{\arcmin}18{\farcs}4)$, corresponding to galactic coordinates $(\ell,b)=(2.71,-3.09)$. Dense observations were obtained through the standard $I$ filter with a cadence of three times per hour, and occasional observations were made in standard $V$-band. This event was also identified using the MOA collaboration's 1.8m telescope at Mt John Observatory in New Zealand, and alerted independently as MOA-2016-BLG-223 roughly 4 days later \citep{2001MNRAS.327..868B}. OGLE-2016-BLG-0795 was also independently detected as KMT-2016-BLG-0143 by the Korea Microlensing Telescope Network (KMTNet, \citep{2016JKAS...49...37K,Kim2018b}) using three 1.6-m telescopes in CTIO (Chile), SAAO (South Africa) and SSO (Australia). The event was located in two overlapping KMTNet fields BLG03 and BLG43, monitored with a cadence of 15 min. From hereon we will refer to the event only by its OGLE designation. CFHT observations were conducted as has been described in the Survey Design section \ref{survey}.  As this analysis is a demonstration, we only use OGLE, KMT-SSO and our own photometry of the event. Photometry of OGLE and KMTNet was extracted using custom implementations of the difference image analysis \citep{1998ApJ...503..325A}: \cite{Wozniak2000} (OGLE) and \cite{pysis} (KMTNet).

\subsection{\ktcn\ Observations and Photometry}\label{k2phot}

\ktcn\ observed a region of the bulge using the \kepler\ spacecraft's imaging instrument between ${\rm HJD}'$ = 7501 and 7572, where ${\rm HJD}'={\rm HJD}-2450000$. Each observation consisted of a 30 minute integration, composed of 270 shorter exposures with virtually no overheads. Continuous sequences of observations were taken with a 30 minute cadence from ${\rm HJD}' = 7501.1$ to $7527.4$ and from ${\rm HJD}' = 7531.1$ to $7572.4$, in two campaigns dubbed C9a and C9b, with the break used to downlink data to Earth. OGLE-2016-BLG-0795 fell on channel 52 of the \kepler\ detector within the superstamp (see Figure~\ref{fig:fields}), and peaked during C9a. 

Extracting the photometry of \ktcn\ microlensing data is challenging. Since the \ktcn\ superstamp is heavily crowded, each \kepler\ pixel contains dozens of stars, and the pixel response function (PRF) is undersampled. Additionally, with only two reaction wheels, the \kepler\ spacecraft can not maintain stable pointing and drifts by about one pixel every $\sim6.5$ hrs, at which point its thrusters are fired to correct for the drift. This drifting motion imprints strong systematics on measured photometry. If the systematics are caused by either sub-pixel sensitivity variations, or aperture losses, for isolated stars they can usually be detrended to a high degree of accuracy~\citep[e.g.,][]{Vanderburg2014}. However, aperture photomerty techniques do not work well in the \ktcn\ field, but several other methods have shown promise for {\it K2} crowded fields, including PSF photometry~\citep[e.g.,][]{Libralato2016}, difference imaging photometry~\citep[e.g.,][]{2017PASP..129d4501S}, and pixel level decorrelation~\citep[e.g.,][]{Wang2017}.

Here we apply the difference imaging photometry and astrometric detrending methods presented by \citet{2017PASP..129d4501S} and \citet{2015MNRAS.454.4159H}, applied to the calibrated \ktcn\ data. We follow the method of \citet{2017PASP..129j4501Z}, who showed that for \ktcn\ data, microlensing event photometry could be more accurately detrended by simultaneously fitting for the microlensing and detrending parameters, while constraining the microlensed source's flux from ground-based photometry. This method yields robust detections of the microlensing variation of OGLE-2016-BLG-0795, and has been successfully applied to several microlensing events \citep{2017ApJ...849L..31Z,2017PASP..129j4501Z,2018AJ....155...40R}. However, we note that OGLE-2016-BLG-0795 lies ${\sim}7\arcsec$ (${\lesssim}2$ \kepler\ pixels) away from an $I=12.6$ variable star with an amplitude of ${\sim 0.1}$~mag~\citep{2013AcA....63...21S}. It is possible that variations from this star affect the extracted \kt\ photometry for this event, but a full exploration of the associated uncertainties is beyond the scope of this paper.

\subsection{Ground-based data analysis: a short event}\label{ground-based}

A single-lens microlensing event has a \citet{Paczynski1986} lightcurve $A(t)$ with parameters $t_0$, $u_0$ and $\tein$: the time of the maximum magnification, the dimensionless impact parameter and the Einstein radius crossing time (or `timescale'), respectively. The event is observed as a change in flux $F(t)$ at the location of the event
\begin{equation}
    F(t) = F_{\rm s} A(t) + F_{\rm c},
\end{equation}
where $F_{\rm s}$ is the flux of the source star being lensed, and $F_{\rm c}$ accounts for any blended flux that is not lensed. The two linear parameters, $F_{\rm s}$ and $F_{\rm c}$ will be different for each observatory and each filter. 

\begin{figure}[htbp]
    %\resizebox{0.5\textwidth}{!}{\includegraphics{ob0795-lc-groundonly.pdf}}
    \resizebox{0.5\textwidth}{!}{\includegraphics{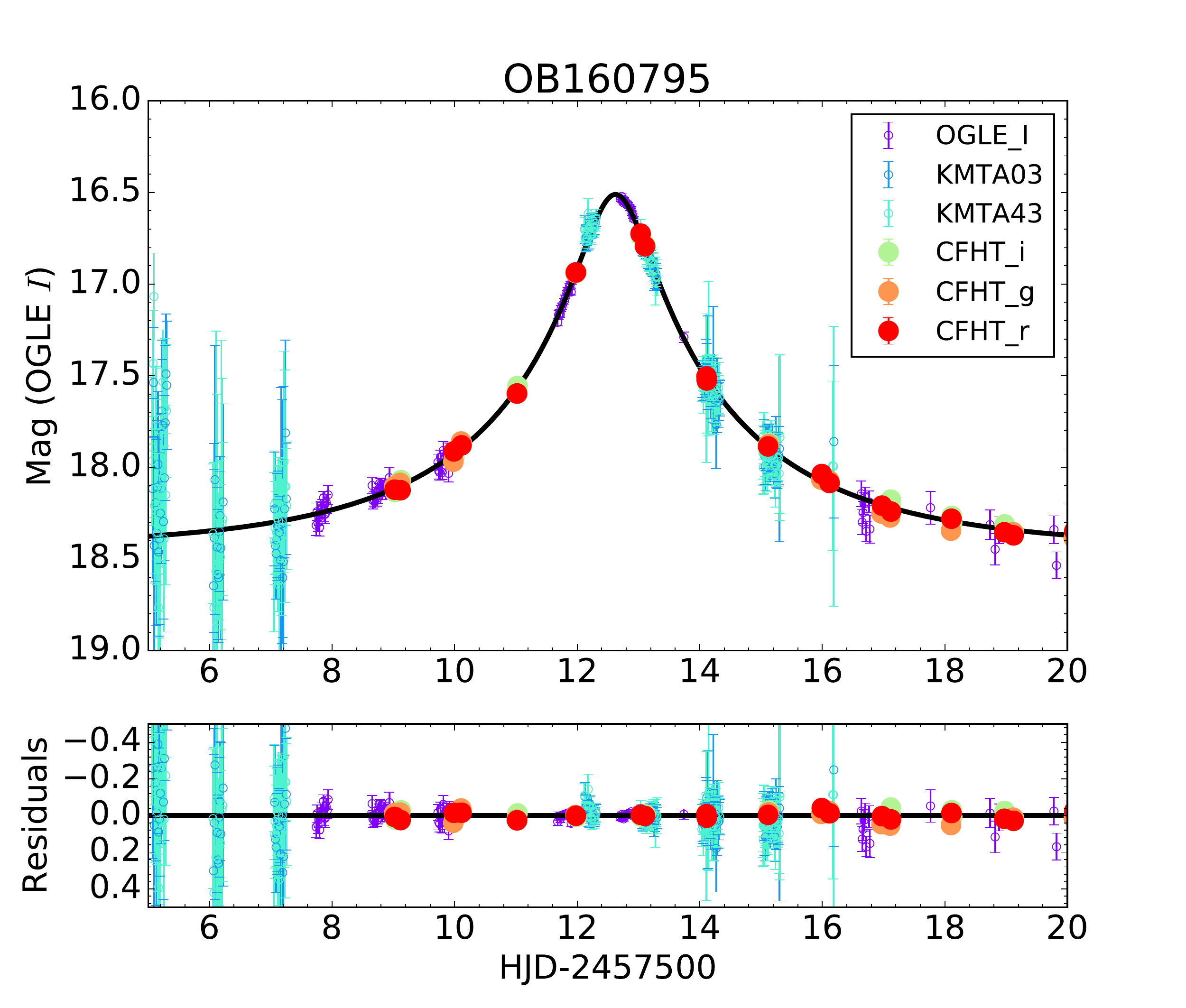}}
    \caption{Ground-based lightcurve of OGLE-2016-BLG-0795. In the top panel, the black line shows the best fit model. The bottom panel shows the residual from the best model.}
    \label{fig:lcground}
\end{figure}

We simultaneously fit a Paczynski lightcurve to the OGLE $I$ band photometry, KMT-SSO $I$ band photometry and our $(g, r, i)_{\rm inst}$ difference photometry using the \texttt{emcee} ensemble sampler \citep{2013PASP..125..306F}. The event has a short timescale of $\tein=4.40 \pm 0.10$ day. The best fit parameters of the model are shown in the first data column of Table~\ref{table:parm}.

\subsection{Constraints on $(K_{p} - r_{\rm PS1})_{\rm s}$ from CFHT data}\label{constrain}

Using model independent regression we find $F_{{\rm s}, r}/F_{{\rm s}, i} = 1.017 \pm 0.008$, which we convert to an instrumental color $(r - i)_{\rm inst} = m_{0,r} - m_{0,i} - 2.5\log(F_{{\rm s}, r}/F_{{\rm s}, i}) = 0.461 \pm 0.009$, where $m_{0,r, \rm inst}$ and $m_{0,i, \rm inst}$ are instrumental zeropoints of the $r$ and $i$ bands respectively.\footnote{The instrumental zeropoints for a given field and chip can be found by entering flux values of 1 into \texttt{calibrate\_flux.py}, e.g., \texttt{python calibrate\_flux.py -e OB160795 1 0 1 0 1 0}.} Using the photometric transformations derived from Section~\ref{PanSTARRS}, we obtain $(r - i)_{\rm PS1} = 0.54 \pm 0.03$, which includes systematic uncertainties in the photometric calibration.  According to \cite{2013ApJ...769...88N}, the extinction parameters of this event are $A_I$ = 1.04, $R_I$ = 1.18. After applying the color-color relation given in Table~\ref{table:kp}, we obtain that the source $(K_{p} - r_{\rm PS1})_{\rm s}$ color is $0.02 \pm 0.02$.

\begin{figure}[htbp]
    \resizebox{0.5\textwidth}{!}{\includegraphics{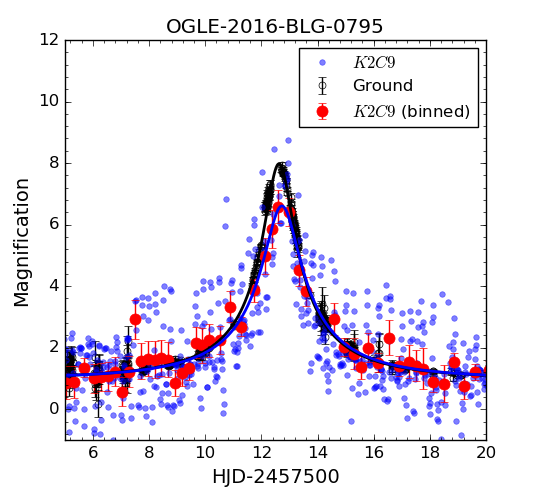}}
    \caption{Lightcurves of event OGLE-2016-BLG-0795 in views of ground-based data (OGLE + CFHT + KMT-SSO, black circles) and \kepler\ data. For \kepler\ data, the blue dots are original data and the red dots are binned data according to trend. The black and blue solid lines are best-fit theoretical light curves for ground-based and \kepler\ observations.}
    \label{fig:lc}
\end{figure}

\subsection{Measuring Parallax from {\it K2} and Ground-based Data}

Armed with an estimate for $K_p$, we then simultaneously modeled the \ktcn\ and ground-based data using the ``detrend+$\mu$lensing model" given by \citet{2017PASP..129j4501Z}. We show the best fitting model with the ground-based and \ktcn\ data in Figure~\ref{fig:lc}. The detrending reveals a clear microlensing signal in the \ktcn\ data with a peak time and magnification similar to those of the ground based lightcurve. The space-based microlensing parallax parameter can be approximated as \citep{1994ApJ...421L..75G}
\begin{equation}
    \pievec \approx \frac{\mathrm{au}}{D_\perp}\left(\frac{\Delta{t_0}}{\tein}, \Delta{u_{0}}\right),
\end{equation}
where $\Delta{t_0}$ and $\Delta{u_{0}}$ are the differences in event peak time $t_0$ and impact parameter $u_0$ as seen from the \kepler\ satellite and Earth, and $D_\perp$ is the projected separation between \kepler\ and the Earth at the time of the event. Since only the absolute value of $|u_0|$ can be measured from the light curves, we have a four-fold degeneracy in the microlening parallax measurement \citep{1966MNRAS.134..315R,1994ApJ...421L..75G}. We summarize the modeling results in Table \ref{table:parm}.

The similarity of $|\uzearth|$ and $|\uzkep|$ leads to two very different possible values of the magnitude of the parallax, $\pie{\sim}0.97$ for the $(+,-)$ and $(-,+)$ solutions where the source appears to pass on opposite sides of the lens to each observer, or $\pie{\sim}0.13$ for the $(+,+)$ and $(-,-)$ solutions, where the source appears to pass on the same side. The Rich argument can be used to probabilistically distinguish between the two scenarios~\citep[see][for full details]{2015ApJ...804...20C}. Succinctly, the Rich argument states that we can expect $|\Delta\uzero|$ to be of the same order of magnitude as $|\Delta\tzero/\tein|$. Therefore, for an event where they are the same order of magnitude, a large-parallax solution requires fine tuning of the source trajectory angle to match the amplitudes but not the signs of $\uzearth$ and $\uzkep$ to within ${\sim}|\Delta \tzero|/\tein$. Quantitatively, the probability of this occurring is ${\sim}(\pi_{{\rm E}+}/\pi_{{\rm E}-})^2$, where $\pi_{{\rm E}+}$ is the small-parallax solution, and $\pi_{{\rm E}-}$ is the large parallax solution. So, for OGLE-2016-BLG-0795 the probability of the large parallax solution being correct is only ${\sim}2$\%.

\subsection{Properties of the Source}

From the dereddened color and magnitude of the source star we can measure its angular radius $\theta_*$. Had we measured finite source effects, we could combine these with the $\theta_*$ to measure the angular Einstein ring radius. Even without a measurement of finite source effects, we can use their absence to place limits on the angular Einstein radius.

\begin{figure}
    \centering
    \includegraphics[width=\columnwidth]{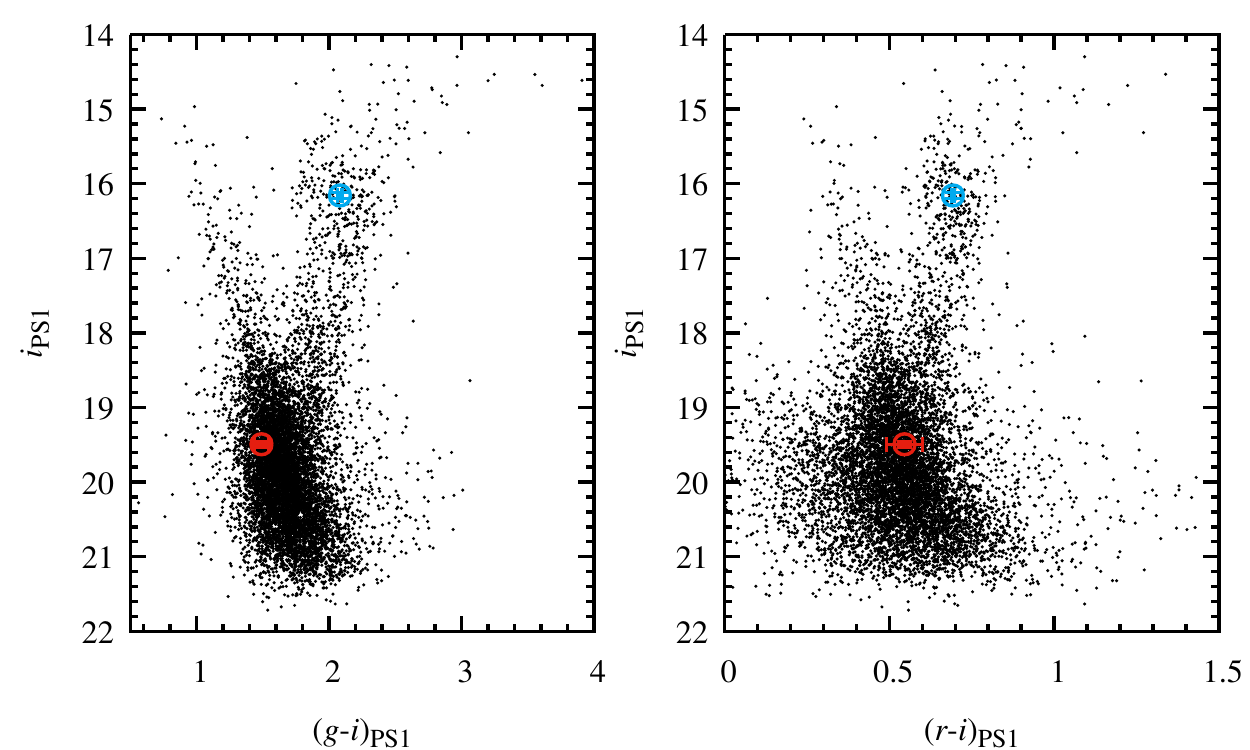}
    \caption{Calibrated color-magnitude diagrams of a $90''$ radius region around OGLE-2016-BLG-0795 ($i$ versus $g-i$ on the left, and $i$ versus $r-i$ on the right). The red point indicates the position of the source, and the blue point shows the centroid of the red clump.}
    \label{fig:cmd}
\end{figure}

To deredden the source we measure the extinction in the $g, r$ and $i$ bands by measuring the red clump centroid in the color magnitude diagrams shown in Figure~\ref{fig:cmd}. We estimate the red clump magnitude to be $i_{\rm cl,PS1}=16.16\pm0.05$, and its colors $(g-i)_{\rm cl,PS1}=2.105\pm0.015$ and $(r-i)_{\rm cl,PS1}=0.696\pm0.004$. Using estimates of the dereddened clump magnitude from \citet{2016MNRAS.456.2692N} and the transformations of \citet{2016ApJ...822...66F} we compute the intrinsic magnitude and colors of the clump to be $i_{\rm cl,PS1,0}=14.836\pm0.05$, $(g-i)_{\rm cl,PS1,0}=1.01\pm0.03$, and $(r-i)_{\rm cl,PS1,0}=0.23\pm0.03$. Subtracting these we find the extinction and reddening to be $A_i=1.32\pm0.07$, $E(g-i)=1.10\pm0.04$, and $E(r-i)=0.46\pm0.03$. With these values, we can deredden the source magnitudes and colors to $i_{\rm s,PS1,0}=18.13\pm 0.08$, $(g-i)_{\rm s,PS1,0}=0.36\pm0.05$, and $(r-i)_{\rm s,PS1,0}=0.08\pm0.04$. Using the color-angular diameter relations from equations~\ref{eq:ad}, \ref{eq:zmld} and Table~\ref{table:angular}, we measure the source radius to be $\theta_*=0.86\pm0.05$~$\mu$as using the $(g-i)$ relation and $\theta_*=0.87\pm0.07$~$\mu$as using the $(r-i)$ relation.

We do not measure finite source effects in the lightcurve of OGLE-2016-BLG-0795, but we can take the impact parameter as a crude upper limit on the ratio of the angular source radius to the angular Einstein radius. This allows us to set a lower limit on $\thetae$,
\begin{equation}
\thetae > \frac{\theta_*}{|u_0|} = 0.0067\pm 0.0006~\text{mas},
\end{equation}
where we have used the more accurate of the source angular radius from $(g-i)$ relation and the largest value of $\uzero$ from all of the event models and observatories. 

\subsection{An Estimate the Mass and Distance of the Lens}

We can combine our measurement of the parallax and our limit on the angular Einstein radius to place limits on the mass and distance of the lens. From equation~\ref{eq:mass} we place lower limits on the mass of the lens $M_{\rm L}\gtrsim 0.8 M_{\rm Jup}$ for the large parallax solutions and $M_{\rm L}\gtrsim 7 M_{\rm Jup}$ for the more likely small parallax solutions. The most probable solution therefore is not likely to be a free-floating planet, though we can not rule out a massive free-floating planet with a mass near the planet-brown dwarf boundary.

\begin{figure*}[t!]
    \centering
    \begin{subfigure}
    \centering
    \includegraphics[width=\columnwidth]{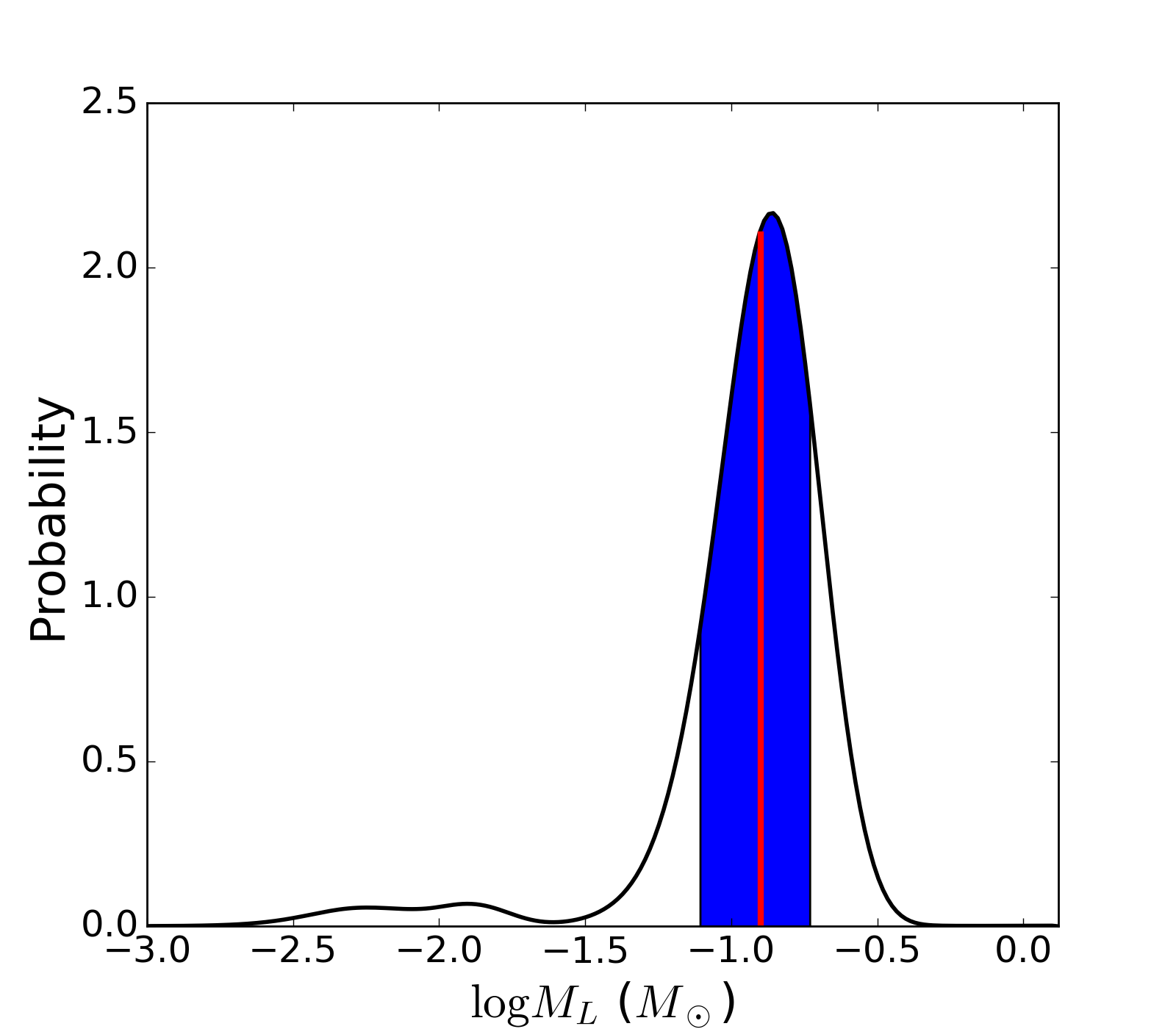}
    \end{subfigure}
    ~
    \begin{subfigure}
    \centering
    \includegraphics[width=\columnwidth]{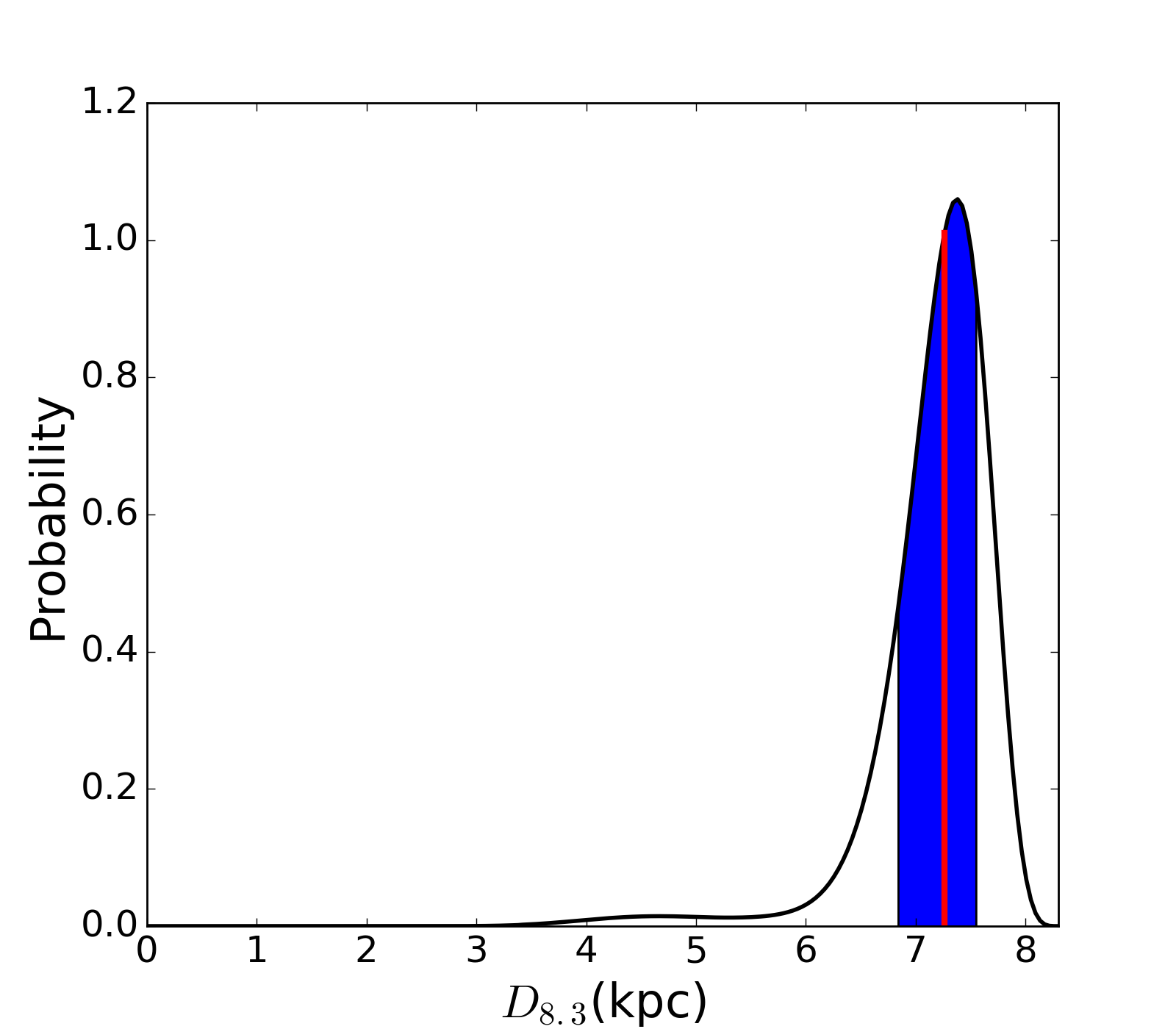}
    \end{subfigure}

    \caption{Posterior probabilities of mass and distance of the lens OGLE-2016-BLG-0795L. In each panel, the red solid vertical line represents the median value and the blue region is the 1$\sigma$ range of the distribution.}
    \label{fig:massdis}
\end{figure*}

We can improve on our lower mass limits by performing a Bayesian analysis. To estimate the mass and distance of the lens of OGLE-2016-BLG-0795, we applied the Galactic model and Bayesian model described in \cite{Zhu2017}. The resulting posterior distributions are shown in Figure~\ref{fig:massdis}, which account for both possible parallax solutions. We find that the most probable lens mass is $M_{\rm L} = 0.13_{-0.05}^{+0.06} M_\odot$ and the most probable lens distance is $D_{\rm L} = 7.26_{-0.46}^{+0.33}$ kpc (assuming the source distance is 8.3 kpc). These values strongly suggest that the lens is a low-mass star in the Galactic bulge, though there is a low probability tail down to ${\sim}3 {\rm M_{Jup}}$ and distances consistent with the Galactic disk.

\begin{table*}[!htb]
    \centering
    \caption{Best-fit parameters and 1-sigma uncertainties of 4-fold parallax degenerate fits to OGLE-2016-BLG-0795.}
    \begin{tabular}{c|c|c|c|c|c}
    \hline
    \hline
    Parameters & OGLE+CFHT & OGLE+CFHT+{\it K2} & OGLE+CFHT+{\it K2} & OGLE+CFHT+{\it K2} & OGLE+CFHT+{\it K2} \\
            & & (+,-) & (-,+) & (+,+) & (-,-) \\
    \hline
    $t_{0,Earth}-2450000$ (d) & 7512.624(4) & 7512.630(4) & 7512.632(4) & 7512.628(4) & 7512.630(4) \\
    $u_0$  & 0.127(4) & 0.128(4) & -0.125(4) & 0.126(4) & -0.127(4) \\
    $t_E$ (d) & 4.40(9) & 4.54(10) & 4.54(11) & 4.43(10) & 4.51(10) \\
    $\pi_{E,N}$ & ... & -0.961(39) & 0.916(15) & 0.072(23) & -0.109(24) \\
    $\pi_{E,E}$ & ... & 0.099(29) & -0.294(30) & -0.097(23) & -0.063(29) \\
    $\pie$ & ... & 0.97(4) & 0.96(2) & 0.12(3) & 0.13(3) \\
    $(K_p - r)_S$ & ... & 0.034(17) & 0.033(30) & 0.042(15) & 0.044(17) \\
    $I({\rm OGLE})$ & 18.81(5) & 18.80(5) & 18.83(6) & 18.81(6) & 18.81(6) \\
    $g_{\rm s,PS1}$ & 20.91(5) & 20.92(6) & 20.98(6) & 20.98(6) & 21.00(6) \\
    $r_{\rm s,PS1}$ & 19.99(5) & 19.97(5) & 20.04(6) & 20.02(6) & 20.05(6) \\
    $i_{\rm s,PS1}$ & 19.45(5) & 19.43(5) & 19.49(6) & 19.48(6) & 19.51(6) \\
    $\chi^2$     & ... & 2502.7 & 2502.2 & 2502.5 & 2502.5 \\
    \hline
    \end{tabular}\\
    {{Note}---For the four-fold parallax degeneracy fits, we require $(K_p - r_{\rm PS1})_S = 0.02 \pm 0.02$ and allow a 3-sigma range. See Equation 9 of \cite{Zhu2017} for the definition of the sign of the 4 solutions.}
    
    \label{table:parm}
\end{table*}

\section{Discussion}\label{discussion}

One of the principal goals of our survey was to aid the measurement of free-floating planet parallaxes with \ktcn. The recent limits on the occurrence rate of free-floating or wide-orbit planets by \citet{Mroz2017} reduced the expected number of events in the \ktcn\ superstamp to less than one during the campaign. However, the discovery of a roughly Neptune-mass free-floating planet \emph{candidate} in the event OGLE-2016-BLG-1540~\citep{2018AJ....155..121M}, which occurred in the \ktcn\ superstamp ${\sim}34$ days after the campaign ended shows that the search was not hopeless. We measured significantly magnified data on ${\sim} 10$ nights of the ${\sim}4$~d short-timescale event OGLE-2016-BLG-0795 that we have analyzed, demonstrating that our data could be used to measure the source color of ${\sim}1$~day timescale events caused by free-floating planet candidates. No such candidate event has been recognized as occurring during \ktcn\ yet, but it is possible that such an event has not yet been noticed. Weather conditions were generally poor during the campaign, especially in Chile, likely due to an El Ni{\~ n}o event, so it is possible that a search for short timescale events in our data set could yield new events. Even if no free-floating planet candidate events are found to have occurred during \ktcn\, our data set provides valuable source color information for $\nevents$ longer timescale microlensing events that will be a valuable contribution to the sample of events being used to measure the relative planet abundance between the disk and bulge.

The 2016 \mcms\ is the first optical microlensing survey conducted toward the Galactic bulge from Hawaii. Hawaii's northern latitude, $+19\degr$, presents several problems to microlensing observations towards the Galactic bulge at declinations of ${\sim}-30\degr$, including a small range of observable hour angles, observations at high airmass, and a shorter season of observations. Despite these problems, there are also advantages, including excellent seeing that is of huge importance to crowded field observations, a high probability of good weather, and a longitude offset from the best southern hemisphere sites. While Hawaii is at a similar longitude to New Zealand (host to MOA) and Australia (host to a KMTNet node), the sites are almost perfectly complementary. Both Mt John and Siding Spring suffer from relatively high probabilities of poor weather, and significantly worse seeing than Hawaii. This means that at $I{\sim}18$ one of our 20~s CFHT $i$-band exposures has the same statistical power as ${\sim}9$ survey exposures from KMTNet-SSO or MOA,\footnote{Based on a comparison of the unweighted sigma-clipped standard deviation our aligned baseline data for OGLE-2016-BLG-0795 (0.036 mag) to the median normalized uncertainty of sigma-clipped KMTNet-SSO field 43 data (0.126 mag).} thanks to the better seeing. Note though that this assessment ignores the value of cadence and coverage, both of which are more challenging for CFHT to provide as it is not dedicated to microlensing observations. The wide-field optical and infra-red telescopes on Mauna Kea (CFHT-MegaCam, CFHT-WIRCAM, Subaru-HyperSuprimeCam, and UKIRT-WFCAM) can therefore add significant value to the existing microlensing survey networks. This has been demonstrated by our data set's contribution to several planet discoveries in 2016~\citep{2017AJ....154....3K,2018AJ....155...40R,2017AJ....154..223H}.

\section{Conclusion}\label{conclusion}

We have presented the the analysis and first data release of the 2016 CFHT-\ktcn\ Multi-color Microlensing Survey. The data release contains $g,r,i$ difference imaging photometry lightcurves of all previously identified microlensing events in our fields, and PSF photometry of our entire fields calibrated to PanSTARRS DR1. We have derived color-angular diameter relations in the PanSTARRS bandpasses, and a color-color relation to derive magnitudes in the \kepler\ bandpass from our photometry. To demonstrate the use of our data, we analyzed the short-timescale microlensing event OGLE-2016-BLG-0795 and measure its microlensing parallax. We find that for the most likely solution of the event, the lens is a low-mass star in the Galactic bulge.

\vspace{6pt}

We thank Yossi Shvartzvald, Andrew Gould, Radek Poleski, Szymon Koz{\l}owski, Yoon-Hyun Ryu, Simon Prunet and Arthur Adams for discussions. We thank the MOA collaboration for making their event locations publicly available. This research uses data obtained through the Telescope Access Program (TAP), which has been funded by the National Astronomical Observatories of China, the Chinese Academy of Sciences (the Strategic Priority Research Program "The Emergence of Cosmological Structures" Grant No. XDB09000000), and the Special Fund for Astronomy from the Ministry of Finance. This work was partly supported by the National Science Foundation of China (Grant No. 11333003, 11390372 to SM). This work was performed in part under contract with the California Institute of Technology (Caltech)/Jet Propulsion Laboratory (JPL) funded by NASA through the Sagan Fellowship Program executed by the NASA Exoplanet Science Institute. Work by MTP was partially supported by NASA grants NNX16AC62G and NNG16PJ32C. Based on observations obtained with MegaPrime/MegaCam, a joint project of CFHT and CEA/DAPNIA, at the Canada-France-Hawaii Telescope (CFHT) which is operated by the National Research Council (NRC) of Canada, the Institut National des Science de l'Univers of the Centre National de la Recherche Scientifique (CNRS) of France, and the University of Hawaii. The authors wish to recognize and acknowledge the very significant cultural role and reverence that the summit of Mauna Kea has always had within the indigenous Hawaiian community.  We are most fortunate to have the opportunity to conduct observations from this mountain. The OGLE project has received funding from the National Science Centre, Poland, grant MAESTRO 2014/14/A/ST9/00121 to AU. This research has made use of the KMTNet system operated by the Korea Astronomy and Space Science Institute (KASI) and the data were obtained at three host sites of CTIO in Chile, SAAO in South Africa, and SSO in Australia. Work by KHH was support by KASI grant 2017-1-830-03. The Pan-STARRS1 Surveys (PS1) and the PS1 public science archive have been made possible through contributions by the Institute for Astronomy, the University of Hawaii, the Pan-STARRS Project Office, the Max-Planck Society and its participating institutes, the Max Planck Institute for Astronomy, Heidelberg and the Max Planck Institute for Extraterrestrial Physics, Garching, The Johns Hopkins University, Durham University, the University of Edinburgh, the Queen's University Belfast, the Harvard-Smithsonian Center for Astrophysics, the Las Cumbres Observatory Global Telescope Network Incorporated, the National Central University of Taiwan, the Space Telescope Science Institute, the National Aeronautics and Space Administration under Grant No. NNX08AR22G issued through the Planetary Science Division of the NASA Science Mission Directorate, the National Science Foundation Grant No. AST-1238877, the University of Maryland, Eotvos Lorand University (ELTE), the Los Alamos National Laboratory, and the Gordon and Betty Moore Foundation.

\software{FITSH: Software Package for Image Processing ascl:1111.014 \citep{2012MNRAS.421.1825P}, ISIS: A method for optimal image subtraction ascl:9909.003 \citep{1998ApJ...503..325A,2000A&AS..144..363A}, WCSTools: Image Astrometry Toolkit ascl:1109.015 \citep{Mink2002}, emcee \citep{2013PASP..125..306F}, astropy \citep{Astropy2013}, numpy, scipy, matplotlib, gnuplot}

\bibliography{Zang.bib}

\begin{thebibliography}{}
\expandafter\ifx\csname natexlab\endcsname\relax\def\natexlab#1{#1}\fi

\bibitem[{{Adams} {et~al.}(2018){Adams}, {Boyajian}, \& {von
  Braun}}]{Adams2018}
{Adams}, A.~D., {Boyajian}, T.~S., \& {von Braun}, K. 2018, \mnras, 473, 3608

\bibitem[{{Alard}(2000)}]{2000A&AS..144..363A}
{Alard}, C. 2000, \aaps, 144, 363

\bibitem[{{Alard} \& {Lupton}(1998)}]{1998ApJ...503..325A}
{Alard}, C., \& {Lupton}, R.~H. 1998, \apj, 503, 325

\bibitem[{{Albrow} {et~al.}(2009){Albrow}, {Horne}, {Bramich}, {Fouqu{\'e}},
  {Miller}, {Beaulieu}, {Coutures}, {Menzies}, {Williams}, {Batista},
  {Bennett}, {Brillant}, {Cassan}, {Dieters}, {Dominis Prester}, {Donatowicz},
  {Greenhill}, {Kains}, {Kane}, {Kubas}, {Marquette}, {Pollard}, {Sahu},
  {Tsapras}, {Wambsganss}, \& {Zub}}]{pysis}
{Albrow}, M.~D., {Horne}, K., {Bramich}, D.~M., {et~al.} 2009, \mnras, 397,
  2099

\bibitem[{{Alcock} {et~al.}(1995){Alcock}, {Allsman}, {Alves}, {Axelrod},
  {Bennett}, {Cook}, {Freeman}, {Griest}, {Guern}, {Lehner}, {Marshall},
  {Peterson}, {Pratt}, {Quinn}, {Rodgers}, {Stubbs}, \&
  {Sutherland}}]{Alcock1995}
{Alcock}, C., {Allsman}, R.~A., {Alves}, D., {et~al.} 1995, \apjl, 454, L125

\bibitem[{{Alonso-Garc{\'{\i}}a} {et~al.}(2012){Alonso-Garc{\'{\i}}a}, {Mateo},
  {Sen}, {Banerjee}, {Catelan}, {Minniti}, \& {von
  Braun}}]{2012AJ....143...70A}
{Alonso-Garc{\'{\i}}a}, J., {Mateo}, M., {Sen}, B., {et~al.} 2012, \aj, 143, 70

\bibitem[{{Astropy Collaboration} {et~al.}(2013){Astropy Collaboration},
  {Robitaille}, {Tollerud}, {Greenfield}, {Droettboom}, {Bray}, {Aldcroft},
  {Davis}, {Ginsburg}, {Price-Whelan}, {Kerzendorf}, {Conley}, {Crighton},
  {Barbary}, {Muna}, {Ferguson}, {Grollier}, {Parikh}, {Nair}, {Unther},
  {Deil}, {Woillez}, {Conseil}, {Kramer}, {Turner}, {Singer}, {Fox}, {Weaver},
  {Zabalza}, {Edwards}, {Azalee Bostroem}, {Burke}, {Casey}, {Crawford},
  {Dencheva}, {Ely}, {Jenness}, {Labrie}, {Lim}, {Pierfederici}, {Pontzen},
  {Ptak}, {Refsdal}, {Servillat}, \& {Streicher}}]{Astropy2013}
{Astropy Collaboration}, {Robitaille}, T.~P., {Tollerud}, E.~J., {et~al.} 2013,
  \aap, 558, A33

\bibitem[{{Bertin} \& {Arnouts}(1996)}]{1996A&AS..117..393B}
{Bertin}, E., \& {Arnouts}, S. 1996, \aaps, 117, 393

\bibitem[{{Bond} {et~al.}(2001){Bond}, {Abe}, {Dodd}, {Hearnshaw}, {Honda},
  {Jugaku}, {Kilmartin}, {Marles}, {Masuda}, {Matsubara}, {Muraki}, {Nakamura},
  {Nankivell}, {Noda}, {Noguchi}, {Ohnishi}, {Rattenbury}, {Reid}, {Saito},
  {Sato}, {Sekiguchi}, {Skuljan}, {Sullivan}, {Sumi}, {Takeuti}, {Watase},
  {Wilkinson}, {Yamada}, {Yanagisawa}, \& {Yock}}]{2001MNRAS.327..868B}
{Bond}, I.~A., {Abe}, F., {Dodd}, R.~J., {et~al.} 2001, \mnras, 327, 868

\bibitem[{{Boulade} {et~al.}(2003){Boulade}, {Charlot}, {Abbon}, {Aune},
  {Borgeaud}, {Carton}, {Carty}, {Da Costa}, {Deschamps}, {Desforge},
  {Eppell{\'e}}, {Gallais}, {Gosset}, {Granelli}, {Gros}, {de Kat}, {Loiseau},
  {Ritou}, {Rouss{\'e}}, {Starzynski}, {Vignal}, \&
  {Vigroux}}]{2003SPIE.4841...72B}
{Boulade}, O., {Charlot}, X., {Abbon}, P., {et~al.} 2003, in \procspie, Vol.
  4841, Instrument Design and Performance for Optical/Infrared Ground-based
  Telescopes, ed. M.~{Iye} \& A.~F.~M. {Moorwood}, 72--81

\bibitem[{{Boyajian} {et~al.}(2014){Boyajian}, {van Belle}, \& {von
  Braun}}]{Boyajian2014}
{Boyajian}, T.~S., {van Belle}, G., \& {von Braun}, K. 2014, \aj, 147, 47

\bibitem[{{Boyajian} {et~al.}(2012){Boyajian}, {von Braun}, {van Belle},
  {McAlister}, {ten Brummelaar}, {Kane}, {Muirhead}, {Jones}, {White},
  {Schaefer}, {Ciardi}, {Henry}, {L{\'o}pez-Morales}, {Ridgway}, {Gies}, {Jao},
  {Rojas-Ayala}, {Parks}, {Sturmann}, {Sturmann}, {Turner}, {Farrington},
  {Goldfinger}, \& {Berger}}]{2012ApJ...757..112B}
{Boyajian}, T.~S., {von Braun}, K., {van Belle}, G., {et~al.} 2012, \apj, 757,
  112

\bibitem[{{Boyajian} {et~al.}(2013){Boyajian}, {von Braun}, {van Belle},
  {Farrington}, {Schaefer}, {Jones}, {White}, {McAlister}, {ten Brummelaar},
  {Ridgway}, {Gies}, {Sturmann}, {Sturmann}, {Turner}, {Goldfinger}, \&
  {Vargas}}]{2013ApJ...771...40B}
---. 2013, \apj, 771, 40

\bibitem[{{Calchi Novati} {et~al.}(2015){Calchi Novati}, {Gould}, {Udalski},
  {Menzies}, {Bond}, {Shvartzvald}, {Street}, {Hundertmark}, {Beichman}, {Yee},
  {Carey}, {Poleski}, {Skowron}, {Koz{\l}owski}, {Mr{\'o}z}, {Pietrukowicz},
  {Pietrzy{\'n}ski}, {Szyma{\'n}ski}, {Soszy{\'n}ski}, {Ulaczyk},
  {Wyrzykowski}, {OGLE Collaboration}, {Albrow}, {Beaulieu}, {Caldwell},
  {Cassan}, {Coutures}, {Danielski}, {Dominis Prester}, {Donatowicz}, {Lon{\v
  c}ari{\'c}}, {McDougall}, {Morales}, {Ranc}, {Zhu}, {PLANET Collaboration},
  {Abe}, {Barry}, {Bennett}, {Bhattacharya}, {Fukunaga}, {Inayama},
  {Koshimoto}, {Namba}, {Sumi}, {Suzuki}, {Tristram}, {Wakiyama}, {Yonehara},
  {MOA Collaboration}, {Maoz}, {Kaspi}, {Friedmann}, {Wise Group}, {Bachelet},
  {Figuera Jaimes}, {Bramich}, {Tsapras}, {Horne}, {Snodgrass}, {Wambsganss},
  {Steele}, {Kains}, {RoboNet Collaboration}, {Bozza}, {Dominik},
  {J{\o}rgensen}, {Alsubai}, {Ciceri}, {D'Ago}, {Haugb{\o}lle}, {Hessman},
  {Hinse}, {Juncher}, {Korhonen}, {Mancini}, {Popovas}, {Rabus}, {Rahvar},
  {Scarpetta}, {Schmidt}, {Skottfelt}, {Southworth}, {Starkey}, {Surdej},
  {Wertz}, {Zarucki}, {MiNDSTEp Consortium}, {Gaudi}, {Pogge}, {DePoy}, \&
  {{$\mu$}FUN Collaboration}}]{2015ApJ...804...20C}
{Calchi Novati}, S., {Gould}, A., {Udalski}, A., {et~al.} 2015, \apj, 804, 20

\bibitem[{{Clanton} \& {Gaudi}(2017)}]{2017ApJ...834...46C}
{Clanton}, C., \& {Gaudi}, B.~S. 2017, \apj, 834, 46

\bibitem[{{Dong} {et~al.}(2007){Dong}, {Udalski}, {Gould}, {Reach}, {Christie},
  {Boden}, {Bennett}, {Fazio}, {Griest}, {Szyma{\'n}ski}, {Kubiak},
  {Soszy{\'n}ski}, {Pietrzy{\'n}ski}, {Szewczyk}, {Wyrzykowski}, {Ulaczyk},
  {Wieckowski}, {Paczy{\'n}ski}, {DePoy}, {Pogge}, {Preston}, {Thompson}, \&
  {Patten}}]{2007ApJ...664..862D}
{Dong}, S., {Udalski}, A., {Gould}, A., {et~al.} 2007, \apj, 664, 862

\bibitem[{{Finkbeiner} {et~al.}(2016){Finkbeiner}, {Schlafly}, {Schlegel},
  {Padmanabhan}, {Juri{\'c}}, {Burgett}, {Chambers}, {Denneau}, {Draper},
  {Flewelling}, {Hodapp}, {Kaiser}, {Magnier}, {Metcalfe}, {Morgan}, {Price},
  {Stubbs}, \& {Tonry}}]{2016ApJ...822...66F}
{Finkbeiner}, D.~P., {Schlafly}, E.~F., {Schlegel}, D.~J., {et~al.} 2016, \apj,
  822, 66

\bibitem[{{Flewelling} {et~al.}(2016){Flewelling}, {Magnier}, {Chambers},
  {Heasley}, {Holmberg}, {Huber}, {Sweeney}, {Waters}, {Chen}, {Farrow},
  {Hasinger}, {Henderson}, {Long}, {Metcalfe}, {Nieto-Santisteban}, {Norberg},
  {Saglia}, {Szalay}, {Rest}, {Thakar}, {Tonry}, {Valenti}, {Werner}, {White},
  {Denneau}, {Draper}, {Hodapp}, {Jedicke}, {Kaiser}, {Kudritzki}, {Price},
  {Wainscoat}, {Chastel}, {McClean}, {Postman}, \&
  {Shiao}}]{2016arXiv161205243F}
{Flewelling}, H.~A., {Magnier}, E.~A., {Chambers}, K.~C., {et~al.} 2016, ArXiv
  e-prints, arXiv:1612.05243

\bibitem[{{Foreman-Mackey} {et~al.}(2013){Foreman-Mackey}, {Hogg}, {Lang}, \&
  {Goodman}}]{2013PASP..125..306F}
{Foreman-Mackey}, D., {Hogg}, D.~W., {Lang}, D., \& {Goodman}, J. 2013, \pasp,
  125, 306

\bibitem[{{Gould}(1992)}]{1992ApJ...392..442G}
{Gould}, A. 1992, \apj, 392, 442

\bibitem[{{Gould}(1994)}]{1994ApJ...421L..75G}
---. 1994, \apjl, 421, L75

\bibitem[{{Gould}(2000)}]{Gould2000}
---. 2000, \apj, 542, 785

\bibitem[{{Gould}(2016)}]{Gould2016}
---. 2016, Journal of Korean Astronomical Society, 49, 123

\bibitem[{{Gould} {et~al.}(2009){Gould}, {Udalski}, {Monard}, {Horne}, {Dong},
  {Miyake}, {Sahu}, {Bennett}, {Wyrzykowski}, {Soszy{\'n}ski}, {Szyma{\'n}ski},
  {Kubiak}, {Pietrzy{\'n}ski}, {Szewczyk}, {Ulaczyk}, {OGLE Collaboration},
  {Allen}, {Christie}, {DePoy}, {Gaudi}, {Han}, {Lee}, {McCormick}, {Natusch},
  {Park}, {Pogge}, {{$\mu$}FUN Collaboration}, {Allan}, {Bode}, {Bramich},
  {Burgdorf}, {Dominik}, {Fraser}, {Kerins}, {Mottram}, {Snodgrass}, {Steele},
  {Street}, {Tsapras}, {RoboNet Collaboration}, {Abe}, {Bond}, {Botzler},
  {Fukui}, {Furusawa}, {Hearnshaw}, {Itow}, {Kamiya}, {Kilmartin}, {Korpela},
  {Lin}, {Ling}, {Masuda}, {Matsubara}, {Muraki}, {Nagaya}, {Ohnishi},
  {Okumura}, {Perrott}, {Rattenbury}, {Saito}, {Sako}, {Skuljan}, {Sullivan},
  {Sumi}, {Sweatman}, {Tristram}, {Yock}, {MOA Collaboration}, {Albrow},
  {Beaulieu}, {Coutures}, {Calitz}, {Caldwell}, {Fouque}, {Martin}, {Williams},
  \& {PLANET Collaboration}}]{Andy07244}
{Gould}, A., {Udalski}, A., {Monard}, B., {et~al.} 2009, \apjl, 698, L147

\bibitem[{{Han} \& {Gould}(1995)}]{1995ApJ...447...53H}
{Han}, C., \& {Gould}, A. 1995, \apj, 447, 53

\bibitem[{{Han} {et~al.}(2017){Han}, {Udalski}, {Gould}, {Lee}, {Shvartzvald},
  {Zang}, {Mao}, {Koz{\l}owski}, {Albrow}, {Chung}, {Hwang}, {Jung}, {Kim},
  {Kim}, {Ryu}, {Shin}, {Yee}, {Zhu}, {Cha}, {Kim}, {Kim}, {Lee}, {Park},
  {KMTNet Collaboration}, {Skowron}, {Mr{\'o}z}, {Pietrukowicz}, {Poleski},
  {Szyma{\'n}ski}, {Soszy{\'n}ski}, {Ulaczyk}, {Pawlak}, {The OGLE
  Collaboration}, {Beichman}, {Bryden}, {Calchi Novati}, {Gaudi}, {Henderson},
  {Howell}, {Jacklin}, {The UKIRT Microlensing Team}, {Penny}, {Fouqu{\'e}},
  {Wang}, \& {CFHT-K2C9 Microlensing Collaboration}}]{2017AJ....154..223H}
{Han}, C., {Udalski}, A., {Gould}, A., {et~al.} 2017, \aj, 154, 223

\bibitem[{{Henderson} \& {Shvartzvald}(2016)}]{Henderson2016b}
{Henderson}, C.~B., \& {Shvartzvald}, Y. 2016, \aj, 152, 96

\bibitem[{{Henderson} {et~al.}(2016){Henderson}, {Poleski}, {Penny}, {Street},
  {Bennett}, {Hogg}, {Gaudi}, {K2 Campaign 9 Microlensing Science Team}, {Zhu},
  {Barclay}, {Barentsen}, {Howell}, {Mullally}, {Udalski}, {Szyma{\'n}ski},
  {Skowron}, {Mr{\'o}z}, {Koz{\l}owski}, {Wyrzykowski}, {Pietrukowicz},
  {Soszy{\'n}ski}, {Ulaczyk}, {Pawlak}, {OGLE Project}, {Sumi}, {Abe},
  {Asakura}, {Barry}, {Bhattacharya}, {Bond}, {Donachie}, {Freeman}, {Fukui},
  {Hirao}, {Itow}, {Koshimoto}, {Li}, {Ling}, {Masuda}, {Matsubara}, {Muraki},
  {Nagakane}, {Ohnishi}, {Oyokawa}, {Rattenbury}, {Saito}, {Sharan},
  {Sullivan}, {Tristram}, {Yonehara}, {MOA Collaboration}, {Bachelet},
  {Bramich}, {Cassan}, {Dominik}, {Figuera Jaimes}, {Horne}, {Hundertmark},
  {Mao}, {Ranc}, {Schmidt}, {Snodgrass}, {Steele}, {Tsapras}, {Wambsganss},
  {RoboNet Project}, {Bozza}, {Burgdorf}, {J{\o}rgensen}, {Calchi Novati},
  {Ciceri}, {D'Ago}, {Evans}, {Hessman}, {Hinse}, {Husser}, {Mancini},
  {Popovas}, {Rabus}, {Rahvar}, {Scarpetta}, {Skottfelt}, {Southworth},
  {Unda-Sanzana}, {The MiNDSTEp Team}, {Bryson}, {Caldwell}, {Haas}, {Larson},
  {McCalmont}, {Packard}, {Peterson}, {Putnam}, {Reedy}, {Ross}, {Van Cleve},
  {K2C9 Engineering Team}, {Akeson}, {Batista}, {Beaulieu}, {Beichman},
  {Bryden}, {Ciardi}, {Cole}, {Coutures}, {Foreman-Mackey}, {Fouqu{\'e}},
  {Friedmann}, {Gelino}, {Kaspi}, {Kerins}, {Korhonen}, {Lang}, {Lee},
  {Lineweaver}, {Maoz}, {Marquette}, {Mogavero}, {Morales}, {Nataf}, {Pogge},
  {Santerne}, {Shvartzvald}, {Suzuki}, {Tamura}, {Tisserand}, \&
  {Wang}}]{2016PASP..128l4401H}
{Henderson}, C.~B., {Poleski}, R., {Penny}, M., {et~al.} 2016, \pasp, 128,
  124401

\bibitem[{{Huang} {et~al.}(2015){Huang}, {Penev}, {Hartman}, {Bakos}, {Bhatti},
  {Domsa}, \& {de Val-Borro}}]{2015MNRAS.454.4159H}
{Huang}, C.~X., {Penev}, K., {Hartman}, J.~D., {et~al.} 2015, \mnras, 454, 4159

\bibitem[{{Husser} {et~al.}(2013){Husser}, {Wende-von Berg}, {Dreizler},
  {Homeier}, {Reiners}, {Barman}, \& {Hauschildt}}]{2013A&A...553A...6H}
{Husser}, T.-O., {Wende-von Berg}, S., {Dreizler}, S., {et~al.} 2013, \aap,
  553, A6

\bibitem[{{Kervella} \& {Fouqu{\'e}}(2008)}]{Kervella2008}
{Kervella}, P., \& {Fouqu{\'e}}, P. 2008, \aap, 491, 855

\bibitem[{{Kim} {et~al.}(2018{\natexlab{a}}){Kim}, {Kim}, {Hwang}, {Albrow},
  {Chung}, {Gould}, {Han}, {Jung}, {Ryu}, {Shin}, {Yee}, {Zhu}, {Cha}, {Kim},
  {Lee}, {Lee}, {Lee}, {Park}, {Pogge}, \& {The KMTNet
  Collaboration}}]{Kim2018a}
{Kim}, D.-J., {Kim}, H.-W., {Hwang}, K.-H., {et~al.} 2018{\natexlab{a}}, \aj,
  155, 76

\bibitem[{{Kim} {et~al.}(2018{\natexlab{b}}){Kim}, {Hwang}, {Kim}, {Albrow},
  {Cha}, {Chung}, {Gould}, {Han}, {Jung}, {Kim}, {Lee}, {Lee}, {Lee}, {Park},
  {Pogge}, {Ryu}, {Shin}, {Shvartzvald}, {Yee}, {Zang}, \& {Zhu}}]{Kim2018b}
{Kim}, H.-W., {Hwang}, K.-H., {Kim}, D.-J., {et~al.} 2018{\natexlab{b}}, ArXiv
  e-prints, arXiv:1801.08166

\bibitem[{{Kim} {et~al.}(2016){Kim}, {Lee}, {Park}, {Kim}, {Cha}, {Lee}, {Han},
  {Chun}, \& {Yuk}}]{2016JKAS...49...37K}
{Kim}, S.-L., {Lee}, C.-U., {Park}, B.-G., {et~al.} 2016, Journal of Korean
  Astronomical Society, 49, 37

\bibitem[{{Koshimoto} {et~al.}(2017){Koshimoto}, {Shvartzvald}, {Bennett},
  {Penny}, {Hundertmark}, {Bond}, {Zang}, {Henderson}, {Suzuki}, {Rattenbury},
  {Sumi}, {and}, {Abe}, {Asakura}, {Bhattacharya}, {Donachie}, {Evans},
  {Fukui}, {Hirao}, {Itow}, {Li}, {Ling}, {Masuda}, {Matsubara}, {Matsuo},
  {Muraki}, {Nagakane}, {Ohnishi}, {Ranc}, {Saito}, {Sharan}, {Shibai},
  {Sullivan}, {Tristram}, {Yamada}, {Yamada}, {Yonehara}, {MOA Collaboration},
  {Gelino}, {Beichman}, {Beaulieu}, {Marquette}, {Batista}, {Keck Team},
  {Friedmann}, {Hallakoun}, {Kaspi}, {Maoz}, {Wise Group}, {Bryden}, {Calchi
  Novati}, {Howell}, {UKIRT Team}, {Wang}, {Mao}, {Fouqu{\'e}}, {Microlensing
  Survey}, {Korhonen}, {J{\o}rgensen}, {Street}, {Tsapras}, {Dominik},
  {Kerins}, {Cassan}, {Snodgrass}, {Bachelet}, {Bozza}, {Bramich}, \& {VST-K2C9
  Team}}]{2017AJ....154....3K}
{Koshimoto}, N., {Shvartzvald}, Y., {Bennett}, D.~P., {et~al.} 2017, \aj, 154,
  3

\bibitem[{{Libralato} {et~al.}(2016){Libralato}, {Bedin}, {Nardiello}, \&
  {Piotto}}]{Libralato2016}
{Libralato}, M., {Bedin}, L.~R., {Nardiello}, D., \& {Piotto}, G. 2016, \mnras,
  456, 1137

\bibitem[{{Ma} {et~al.}(2016){Ma}, {Mao}, {Ida}, {Zhu}, \&
  {Lin}}]{2016MNRAS.461L.107M}
{Ma}, S., {Mao}, S., {Ida}, S., {Zhu}, W., \& {Lin}, D.~N.~C. 2016, \mnras,
  461, L107

\bibitem[{{Magnier} {et~al.}(2016){Magnier}, {Schlafly}, {Finkbeiner}, {Tonry},
  {Goldman}, {R{\"o}ser}, {Schilbach}, {Chambers}, {Flewelling}, {Huber},
  {Price}, {Sweeney}, {Waters}, {Denneau}, {Draper}, {Hodapp}, {Jedicke},
  {Kudritzki}, {Metcalfe}, {Stubbs}, \& {Wainscoast}}]{2016arXiv161205242M}
{Magnier}, E.~A., {Schlafly}, E.~F., {Finkbeiner}, D.~P., {et~al.} 2016, ArXiv
  e-prints, arXiv:1612.05242

\bibitem[{{Mink}(2002)}]{Mink2002}
{Mink}, D.~J. 2002, in Astronomical Society of the Pacific Conference Series,
  Vol. 281, Astronomical Data Analysis Software and Systems XI, ed. D.~A.
  {Bohlender}, D.~{Durand}, \& T.~H. {Handley}, 169

\bibitem[{{Mink}(2011)}]{Mink2011}
{Mink}, J. 2011, {WCSTools: Image Astrometry Toolkit}, Astrophysics Source Code
  Library, , , ascl:1109.015

\bibitem[{{Mr{\'o}z} {et~al.}(2017){Mr{\'o}z}, {Udalski}, {Skowron}, {Poleski},
  {Koz{\l}owski}, {Szyma{\'n}ski}, {Soszy{\'n}ski}, {Wyrzykowski},
  {Pietrukowicz}, {Ulaczyk}, {Skowron}, \& {Pawlak}}]{Mroz2017}
{Mr{\'o}z}, P., {Udalski}, A., {Skowron}, J., {et~al.} 2017, \nat, 548, 183

\bibitem[{{Mr{\'o}z} {et~al.}(2018){Mr{\'o}z}, {Ryu}, {Skowron}, {Udalski},
  {Gould}, {Szyma{\'n}ski}, {Soszy{\'n}ski}, {Poleski}, {Pietrukowicz},
  {Koz{\l}owski}, {Pawlak}, {Ulaczyk}, {OGLE Collaboration}, {Albrow}, {Chung},
  {Jung}, {Han}, {Hwang}, {Shin}, {Yee}, {Zhu}, {Cha}, {Kim}, {Kim}, {Kim},
  {Lee}, {Lee}, {Lee}, {Park}, {Pogge}, \& {KMTNet
  Collaboration}}]{2018AJ....155..121M}
{Mr{\'o}z}, P., {Ryu}, Y.-H., {Skowron}, J., {et~al.} 2018, \aj, 155, 121

\bibitem[{{Mu{\v z}i{\'c}} {et~al.}(2015){Mu{\v z}i{\'c}}, {Scholz}, {Geers},
  \& {Jayawardhana}}]{2015ApJ...810..159M}
{Mu{\v z}i{\'c}}, K., {Scholz}, A., {Geers}, V.~C., \& {Jayawardhana}, R. 2015,
  \apj, 810, 159

\bibitem[{{Nataf} {et~al.}(2013){Nataf}, {Gould}, {Fouqu{\'e}}, {Gonzalez},
  {Johnson}, {Skowron}, {Udalski}, {Szyma{\'n}ski}, {Kubiak},
  {Pietrzy{\'n}ski}, {Soszy{\'n}ski}, {Ulaczyk}, {Wyrzykowski}, \&
  {Poleski}}]{2013ApJ...769...88N}
{Nataf}, D.~M., {Gould}, A., {Fouqu{\'e}}, P., {et~al.} 2013, \apj, 769, 88

\bibitem[{{Nataf} {et~al.}(2016){Nataf}, {Gonzalez}, {Casagrande}, {Zasowski},
  {Wegg}, {Wolf}, {Kunder}, {Alonso-Garcia}, {Minniti}, {Rejkuba}, {Saito},
  {Valenti}, {Zoccali}, {Poleski}, {Pietrzy{\'n}ski}, {Skowron},
  {Soszy{\'n}ski}, {Szyma{\'n}ski}, {Udalski}, {Ulaczyk}, \&
  {Wyrzykowski}}]{2016MNRAS.456.2692N}
{Nataf}, D.~M., {Gonzalez}, O.~A., {Casagrande}, L., {et~al.} 2016, \mnras,
  456, 2692

\bibitem[{{Nemiroff} \& {Wickramasinghe}(1994)}]{Nemiroff1994}
{Nemiroff}, R.~J., \& {Wickramasinghe}, W.~A.~D.~T. 1994, \apjl, 424, L21

\bibitem[{{Ofek}(2008)}]{2008PASP..120.1128O}
{Ofek}, E.~O. 2008, \pasp, 120, 1128

\bibitem[{{Paczy{\'n}ski}(1986)}]{Paczynski1986}
{Paczy{\'n}ski}, B. 1986, \apj, 304, 1

\bibitem[{{P{\'a}l}(2012)}]{2012MNRAS.421.1825P}
{P{\'a}l}, A. 2012, \mnras, 421, 1825

\bibitem[{{Pe{\~n}a Ram{\'{\i}}rez} {et~al.}(2012){Pe{\~n}a Ram{\'{\i}}rez},
  {B{\'e}jar}, {Zapatero Osorio}, {Petr-Gotzens}, \&
  {Mart{\'{\i}}n}}]{2012ApJ...754...30P}
{Pe{\~n}a Ram{\'{\i}}rez}, K., {B{\'e}jar}, V.~J.~S., {Zapatero Osorio}, M.~R.,
  {Petr-Gotzens}, M.~G., \& {Mart{\'{\i}}n}, E.~L. 2012, \apj, 754, 30

\bibitem[{{Penny} {et~al.}(2017){Penny}, {Rattenbury}, {Gaudi}, \&
  {Kerins}}]{Penny2017}
{Penny}, M.~T., {Rattenbury}, N.~J., {Gaudi}, B.~S., \& {Kerins}, E. 2017, \aj,
  153, 161

\bibitem[{{Pickles} \& {Depagne}(2010)}]{2010PASP..122.1437P}
{Pickles}, A., \& {Depagne}, {\'E}. 2010, \pasp, 122, 1437

\bibitem[{{Poleski}(2016)}]{Poleski2016}
{Poleski}, R. 2016, \mnras, 455, 3656

\bibitem[{{Refsdal}(1966)}]{1966MNRAS.134..315R}
{Refsdal}, S. 1966, \mnras, 134, 315

\bibitem[{{Ryu} {et~al.}(2018){Ryu}, {Yee}, {Udalski}, {Bond}, {Shvartzvald},
  {Zang}, {Figuera Jaimes}, {J{\o}rgensen}, {Zhu}, {Huang}, {Jung}, {Albrow},
  {Chung}, {Gould}, {Han}, {Hwang}, {Shin}, {Cha}, {Kim}, {Kim}, {Kim}, {Lee},
  {Lee}, {Lee}, {Park}, {Pogge}, {KMTNet Collaboration}, {Calchi Novati},
  {Carey}, {Henderson}, {Beichman}, {Gaudi}, {Spitzer team}, {Mr{\'o}z},
  {Poleski}, {Skowron}, {Szyma{\'n}ski}, {Soszy{\'n}ski}, {Koz{\l}owski},
  {Pietrukowicz}, {Ulaczyk}, {Pawlak}, {OGLE Collaboration}, {Abe}, {Asakura},
  {Barry}, {Bennett}, {Bhattacharya}, {Donachie}, {Evans}, {Fukui}, {Hirao},
  {Itow}, {Kawasaki}, {Koshimoto}, {Li}, {Ling}, {Masuda}, {Matsubara},
  {Miyazaki}, {Muraki}, {Nagakane}, {Ohnishi}, {Ranc}, {Rattenbury}, {Saito},
  {Sharan}, {Sullivan}, {Sumi}, {Suzuki}, {Tristram}, {Yamada}, {Yamada},
  {Yonehara}, {MOA Collaboration}, {Bryden}, {Howell}, {Jacklin}, {UKIRT
  Microlensing Team}, {Penny}, {Mao}, {Fouqu{\'e}}, {Wang}, {CFHT-K2C9
  Microlensing Survey group}, {Street}, {Tsapras}, {Hundertmark}, {Bachelet},
  {Dominik}, {Li}, {Cross}, {Cassan}, {Horne}, {Schmidt}, {Wambsganss}, {Ment},
  {Maoz}, {Snodgrass}, {Steele}, {RoboNet Team}, {Bozza}, {Burgdorf}, {Ciceri},
  {D'Ago}, {Evans}, {Hinse}, {Kerins}, {Kokotanekova}, {Longa}, {MacKenzie},
  {Popovas}, {Rabus}, {Rahvar}, {Sajadian}, {Skottfelt}, {Southworth}, {von
  Essen}, \& {MiNDSTEp Team}}]{2018AJ....155...40R}
{Ryu}, Y.-H., {Yee}, J.~C., {Udalski}, A., {et~al.} 2018, \aj, 155, 40

\bibitem[{{Schechter} {et~al.}(1993){Schechter}, {Mateo}, \&
  {Saha}}]{1993PASP..105.1342S}
{Schechter}, P.~L., {Mateo}, M., \& {Saha}, A. 1993, \pasp, 105, 1342

\bibitem[{{Scholz} {et~al.}(2012){Scholz}, {Muzic}, {Geers}, {Bonavita},
  {Jayawardhana}, \& {Tamura}}]{2012ApJ...744....6S}
{Scholz}, A., {Muzic}, K., {Geers}, V., {et~al.} 2012, \apj, 744, 6

\bibitem[{{Siverd} {et~al.}(2012){Siverd}, {Beatty}, {Pepper}, {Eastman},
  {Collins}, {Bieryla}, {Latham}, {Buchhave}, {Jensen}, {Crepp}, {Street},
  {Stassun}, {Gaudi}, {Berlind}, {Calkins}, {DePoy}, {Esquerdo}, {Fulton},
  {F{\H u}r{\'e}sz}, {Geary}, {Gould}, {Hebb}, {Kielkopf}, {Marshall}, {Pogge},
  {Stanek}, {Stefanik}, {Szentgyorgyi}, {Trueblood}, {Trueblood}, {Stutz}, \&
  {van Saders}}]{2012ApJ...761..123S}
{Siverd}, R.~J., {Beatty}, T.~G., {Pepper}, J., {et~al.} 2012, \apj, 761, 123

\bibitem[{{Soares-Furtado} {et~al.}(2017){Soares-Furtado}, {Hartman}, {Bakos},
  {Huang}, {Penev}, \& {Bhatti}}]{2017PASP..129d4501S}
{Soares-Furtado}, M., {Hartman}, J.~D., {Bakos}, G.~{\'A}., {et~al.} 2017,
  \pasp, 129, 044501

\bibitem[{{Soszy{\'n}ski} {et~al.}(2013){Soszy{\'n}ski}, {Udalski},
  {Szyma{\'n}ski}, {Kubiak}, {Pietrzy{\'n}ski}, {Wyrzykowski}, {Ulaczyk},
  {Poleski}, {Koz{\l}owski}, {Pietrukowicz}, \&
  {Skowron}}]{2013AcA....63...21S}
{Soszy{\'n}ski}, I., {Udalski}, A., {Szyma{\'n}ski}, M.~K., {et~al.} 2013,
  \actaa, 63, 21

\bibitem[{{Sumi} {et~al.}(2011){Sumi}, {Kamiya}, {Bennett}, {Bond}, {Abe},
  {Botzler}, {Fukui}, {Furusawa}, {Hearnshaw}, {Itow}, {Kilmartin}, {Korpela},
  {Lin}, {Ling}, {Masuda}, {Matsubara}, {Miyake}, {Motomura}, {Muraki},
  {Nagaya}, {Nakamura}, {Ohnishi}, {Okumura}, {Perrott}, {Rattenbury}, {Saito},
  {Sako}, {Sullivan}, {Sweatman}, {Tristram}, {Udalski}, {Szyma{\'n}ski},
  {Kubiak}, {Pietrzy{\'n}ski}, {Poleski}, {Soszy{\'n}ski}, {Wyrzykowski},
  {Ulaczyk}, \& {Microlensing Observations in Astrophysics (MOA)
  Collaboration}}]{2011Natur.473..349S}
{Sumi}, T., {Kamiya}, K., {Bennett}, D.~P., {et~al.} 2011, \nat, 473, 349

\bibitem[{{Tonry} {et~al.}(2012){Tonry}, {Stubbs}, {Lykke}, {Doherty},
  {Shivvers}, {Burgett}, {Chambers}, {Hodapp}, {Kaiser}, {Kudritzki},
  {Magnier}, {Morgan}, {Price}, \& {Wainscoat}}]{2012ApJ...750...99T}
{Tonry}, J.~L., {Stubbs}, C.~W., {Lykke}, K.~R., {et~al.} 2012, \apj, 750, 99

\bibitem[{{Udalski}(2003)}]{2003AcA....53..291U}
{Udalski}, A. 2003, \actaa, 53, 291

\bibitem[{{Udalski} {et~al.}(1994){Udalski}, {Szymanski}, {Kaluzny}, {Kubiak},
  {Mateo}, {Krzeminski}, \& {Paczynski}}]{1994AcA....44..227U}
{Udalski}, A., {Szymanski}, M., {Kaluzny}, J., {et~al.} 1994, \actaa, 44, 227

\bibitem[{{Udalski} {et~al.}(2015{\natexlab{a}}){Udalski}, {Szyma{\'n}ski}, \&
  {Szyma{\'n}ski}}]{2015AcA....65....1U}
{Udalski}, A., {Szyma{\'n}ski}, M., \& {Szyma{\'n}ski}, G. 2015{\natexlab{a}},
  \actaa, 65, 1

\bibitem[{{Udalski} {et~al.}(2015{\natexlab{b}}){Udalski}, {Yee}, {Gould},
  {Carey}, {Zhu}, {Skowron}, {Koz{\l}owski}, {Poleski}, {Pietrukowicz},
  {Pietrzy{\'n}ski}, {Szyma{\'n}ski}, {Mr{\'o}z}, {Soszy{\'n}ski}, {Ulaczyk},
  {Wyrzykowski}, {Han}, {Calchi Novati}, \& {Pogge}}]{2015ApJ...799..237U}
{Udalski}, A., {Yee}, J.~C., {Gould}, A., {et~al.} 2015{\natexlab{b}}, \apj,
  799, 237

\bibitem[{{Van Cleve} \& {Caldwell}(2016)}]{VanCleve2016}
{Van Cleve}, J.~E., \& {Caldwell}, D.~A. 2016, {Kepler Instrument Handbook},
  Tech. rep.

\bibitem[{{Vanderburg} \& {Johnson}(2014)}]{Vanderburg2014}
{Vanderburg}, A., \& {Johnson}, J.~A. 2014, \pasp, 126, 948

\bibitem[{{Wang} {et~al.}(2017{\natexlab{a}}){Wang}, {Hogg}, {Foreman-Mackey},
  \& {Sch{\"o}lkopf}}]{Wang2017}
{Wang}, D., {Hogg}, D.~W., {Foreman-Mackey}, D., \& {Sch{\"o}lkopf}, B.
  2017{\natexlab{a}}, ArXiv e-prints, arXiv:1710.02428

\bibitem[{{Wang} {et~al.}(2017{\natexlab{b}}){Wang}, {Zhu}, {Mao}, {Bond},
  {Gould}, {Udalski}, {Sumi}, {Bozza}, {Ranc}, {Cassan}, {Yee}, {Han}, {Abe},
  {Asakura}, {Barry}, {Bennett}, {Bhattacharya}, {Donachie}, {Evans}, {Fukui},
  {Hirao}, {Itow}, {Kawasaki}, {Koshimoto}, {Li}, {Ling}, {Masuda},
  {Matsubara}, {Miyazaki}, {Muraki}, {Nagakane}, {Ohnishi}, {Rattenbury},
  {Saito}, {Sharan}, {Shibai}, {Sullivan}, {Suzuki}, {Tristram}, {Yamada},
  {Yonehara}, {MOA Collaboration}, {Koz{\L}owski}, {Mr{\'o}z}, {Pawlak},
  {Pietrukowicz}, {Poleski}, {Skowron}, {Soszy{\'n}ski}, {Szyma{\'n}ski},
  {Ulaczyk}, {OGLE Collaboration}, {Beichman}, {Bryden}, {Calchi Novati},
  {Carey}, {Fausnaugh}, {Gaudi}, {Henderson}, {Shvartzvald}, {Wibking},
  {Spitzer Team}, {Albrow}, {Chung}, {Hwang}, {Jung}, {Ryu}, {Shin}, {Cha},
  {Kim}, {Kim}, {Kim}, {Lee}, {Lee}, {Park}, {Pogge}, {KMTNet Collaboration},
  {Street}, {Tsapras}, {Hundertmark}, {Bachelet}, {Dominik}, {Horne}, {Figuera
  Jaimes}, {Wambsganss}, {Bramich}, {Schmidt}, {Snodgrass}, {Steele},
  {Menzies}, \& {RoboNet Collaboration}}]{2017ApJ...845..129W}
{Wang}, T., {Zhu}, W., {Mao}, S., {et~al.} 2017{\natexlab{b}}, \apj, 845, 129

\bibitem[{{Witt} \& {Mao}(1994)}]{Shude1994}
{Witt}, H.~J., \& {Mao}, S. 1994, \apj, 430, 505

\bibitem[{{Wo{\'z}niak} \& {Paczy{\'n}ski}(1997)}]{Wozniak1997}
{Wo{\'z}niak}, P., \& {Paczy{\'n}ski}, B. 1997, \apj, 487, 55

\bibitem[{{Wozniak}(2000)}]{Wozniak2000}
{Wozniak}, P.~R. 2000, \actaa, 50, 421

\bibitem[{{Yee} {et~al.}(2015){Yee}, {Udalski}, {Calchi Novati}, {Gould},
  {Carey}, {Poleski}, {Gaudi}, {Pogge}, {Skowron}, {Koz{\l}owski}, {Mr{\'o}z},
  {Pietrukowicz}, {Pietrzy{\'n}ski}, {Szyma{\'n}ski}, {Soszy{\'n}ski},
  {Ulaczyk}, \& {Wyrzykowski}}]{2015ApJ...802...76Y}
{Yee}, J.~C., {Udalski}, A., {Calchi Novati}, S., {et~al.} 2015, \apj, 802, 76

\bibitem[{{Zhu} {et~al.}(2015){Zhu}, {Udalski}, {Gould}, {Dominik}, {Bozza},
  {Han}, {Yee}, {Calchi Novati}, {Beichman}, {Carey}, {Poleski}, {Skowron},
  {Koz{\l}owski}, {Mr{\'o}z}, {Pietrukowicz}, {Pietrzy{\'n}ski},
  {Szyma{\'n}ski}, {Soszy{\'n}ski}, {Ulaczyk}, {Wyrzykowski}, {OGLE
  Collaboration}, {Gaudi}, {Pogge}, {DePoy}, {Jung}, {Choi}, {Hwang}, {Shin},
  {Park}, {Jeong}, \& {{$\mu$}FUN Collaboration}}]{2015ApJ...805....8Z}
{Zhu}, W., {Udalski}, A., {Gould}, A., {et~al.} 2015, \apj, 805, 8

\bibitem[{{Zhu} {et~al.}(2017{\natexlab{a}}){Zhu}, {Udalski}, {Huang}, {Calchi
  Novati}, {Sumi}, {Poleski}, {Skowron}, {Mr{\'o}z}, {Szyma{\'n}ski},
  {Soszy{\'n}ski}, {Pietrukowicz}, {Koz{\l}owski}, {Ulaczyk}, {Pawlak}, {OGLE
  Collaboration}, {Beichman}, {Bryden}, {Carey}, {Gaudi}, {Gould}, {Henderson},
  {Shvartzvald}, {Yee}, {Spitzer Team}, {Bond}, {Bennett}, {Suzuki},
  {Rattenbury}, {Koshimoto}, {Abe}, {Asakura}, {Barry}, {Bhattacharya},
  {Donachie}, {Evans}, {Fukui}, {Hirao}, {Itow}, {Kawasaki}, {Li}, {Ling},
  {Masuda}, {Matsubara}, {Miyazaki}, {Munakata}, {Muraki}, {Nagakane},
  {Ohnishi}, {Ranc}, {Saito}, {Sharan}, {Sullivan}, {Tristram}, {Yamada},
  {Yonehara}, \& {MOA Collaboration}}]{2017ApJ...849L..31Z}
{Zhu}, W., {Udalski}, A., {Huang}, C.~X., {et~al.} 2017{\natexlab{a}}, \apjl,
  849, L31

\bibitem[{{Zhu} {et~al.}(2017{\natexlab{b}}){Zhu}, {Huang}, {Udalski},
  {Soares-Furtado}, {Poleski}, {Skowron}, {Mr{\'o}z}, {Szyma{\'n}ski},
  {Soszy{\'n}ski}, {Pietrukowicz}, {Koz{\L}owski}, {Ulaczyk}, \&
  {Pawlak}}]{2017PASP..129j4501Z}
{Zhu}, W., {Huang}, C.~X., {Udalski}, A., {et~al.} 2017{\natexlab{b}}, \pasp,
  129, 104501

\bibitem[{{Zhu} {et~al.}(2017{\natexlab{c}}){Zhu}, {Udalski}, {Calchi Novati},
  {Chung}, {Jung}, {Ryu}, {Shin}, {Gould}, {Lee}, {Albrow}, {Yee}, {Han},
  {Hwang}, {Cha}, {Kim}, {Kim}, {Kim}, {Kim}, {Lee}, {Park}, {Pogge}, {KMTNet
  Collaboration}, {Poleski}, {Mr{\'o}z}, {Pietrukowicz}, {Skowron},
  {Szyma{\'n}ski}, {Koz{\l}owski}, {Ulaczyk}, {Pawlak}, {OGLE Collaboration},
  {Beichman}, {Bryden}, {Carey}, {Fausnaugh}, {Gaudi}, {Henderson},
  {Shvartzvald}, {Wibking}, \& {Spitzer Team}}]{Zhu2017}
{Zhu}, W., {Udalski}, A., {Calchi Novati}, S., {et~al.} 2017{\natexlab{c}},
  \aj, 154, 210

\end{thebibliography}

\end{document}